\documentclass[]{emulateapj}
\usepackage{natbib}
\usepackage{epsfig}

\newcommand{\Mb}{M_{\mathrm{b}}}
\newcommand{\Md}{M_{\mathrm{d}}}
\newcommand{\Ms}{M_{\mathrm{s}}}
\newcommand{\Mdt}{M_{\mathrm{d,thick}}}
\newcommand{\ab}{a_{\mathrm{b}}}
\newcommand{\Sb}{\Sigma_{\mathrm{b}}}
\newcommand{\as}{a_{\mathrm{s}}}
\newcommand{\Ss}{\Sigma_{\mathrm{s}}}
\newcommand{\rd}{R_{\mathrm{d}}}
\newcommand{\Sd}{\Sigma_{\mathrm{d}}}
\newcommand{\rdt}{R_{\mathrm{d,thick}}}
\newcommand{\Sdt}{\Sigma_{\mathrm{d,thick}}}
\newcommand{\cdm}{\mathrm{CDM}}
\newcommand{\lcdm}{\Lambda\cdm}

\newcommand{\Ndm}{N_{\mathrm{dm}}}
\newcommand{\Nstars}{N_{\mathrm{stars}}}
\newcommand{\Ngas}{N_{\mathrm{gas}}}
\newcommand{\fgas}{f_{\mathrm{gas}}}
\newcommand{\gas}{\mathrm{gas}}
\newcommand{\Msun}{M_{\odot}}

\newcommand{\MBH}{M_{\mathrm{BH}}}

\newcommand{\Mstar}{\mathcal{M}_{\star}}
\newcommand{\Mgas}{\mathcal{M}_{\gas}}

\newcommand{\Mvir}{M_{\mathrm{vir}}}
\newcommand{\Vvir}{V_{\mathrm{vir}}}
\newcommand{\Vrot}{V_{\mathrm{rot}}}
\newcommand{\VvirA}{V_{\mathrm{vir,A}}}
\newcommand{\VvirB}{V_{\mathrm{vir,B}}}
\newcommand{\Rvir}{R_{\mathrm{vir}}}
\newcommand{\rinit}{r_{\mathrm{init}}}
\newcommand{\rperi}{r_{\mathrm{peri}}}

\newcommand{\qeos}{q_{\mathrm{EOS}}}

\newcommand{\Rd}{R_{\mathrm{d}}}

\newcommand{\beq}{\begin{equation}}
\newcommand{\eeq}{\end{equation}}
\newcommand{\beqa}{\begin{eqnarray}}
\newcommand{\eeqa}{\end{eqnarray}}

\shorttitle{Merger-Driven Disk Galaxy Formation}
\shortauthors{Robertson et al.}

\begin{document}

\title{A Merger-Driven Scenario for\\Cosmological Disk Galaxy Formation}
\author{Brant Robertson\altaffilmark{1,6},
        James  S. Bullock\altaffilmark{2},
	Thomas J. Cox\altaffilmark{1},\\
      Tiziana Di Matteo\altaffilmark{3},
	Lars Hernquist\altaffilmark{1},
     Volker Springel\altaffilmark{4},
     Naoki Yoshida\altaffilmark{5}}
\altaffiltext{1}{Harvard-Smithsonian Center for Astrophysics, 60 Garden St., Cambridge, MA 02138, USA}
\altaffiltext{2}{Department of Physics \& Astronomy, University of California, Irvine, CA 92697, USA}
\altaffiltext{3}{Carnegie-Mellon University, Department of Physics, 5000 Forbes Ave., Pittsburgh, PA 15213, USA}
\altaffiltext{4}{Max-Planck-Institut f\"ur Astrophysik, Karl-Schwarzschild-Stra\ss e  1, 85740 Garching bei M\"unchen, \\Germany}
\altaffiltext{5}{Department of Physics, Nagoya University, Nagoya 464-8602, Japan}
\altaffiltext{6}{brobertson@cfa.harvard.edu}

\begin{abstract}

The violent hierarchical nature of the $\Lambda$-Cold Dark Matter
cosmology poses serious difficulties for the formation of disk
galaxies.  To help resolve these issues, we describe a new,
merger-driven scenario for the cosmological formation of disk galaxies
at high redshifts that supplements the standard model based on dissipational collapse.
In this picture, large gaseous disks may be produced from high-angular
momentum mergers of systems that are gas-dominated, i.e. $\Mgas/(\Mgas+\Mstar) \gtrsim 0.5$
at the height of the merger.
Pressurization from the
multiphase structure of the interstellar medium 
prevents the complete conversion of gas
into stars during the merger, and if enough gas remains to form a disk,
the remnant eventually resembles a disk galaxy.  We perform numerical
simulations of galaxy mergers to study how supernovae feedback 
strength, supermassive black hole growth and feedback, progenitor gas 
fraction, merger mass-ratio, and orbital geometry
impact the formation of remnant disks.
We find that disks can build
angular momentum through mergers and the degree of rotational
support of the baryons in the merger remnant is primarily related to
feedback processes associated with star formation.
Nearly every simulated gas-rich merger remnant 
contains rapidly-rotating stellar substructure, while 
disk-dominated remnants are restricted to form in mergers that are 
gas-dominated at the time of final coalescence.  Typically,
gas-dominated mergers require extreme progenitor
gas fractions ($\fgas>0.8$).
We also show that 
the formation of
rotationally-supported stellar systems in mergers is not restricted to 
idealized
orbits, and both gas-rich major and minor mergers can produce disk-dominated 
stellar 
remnants.
We suggest that the hierarchical nature of
the $\Lambda$-Cold Dark Matter cosmology and the physics of the
interstellar gas may act
together to form spiral galaxies by building the angular momentum of
disks through early, gas-dominated mergers.  Our proposed scenario may be especially
important for galaxy formation at high redshifts, where gas-dominated
mergers are believed to be more common than in the local Universe.

\end{abstract}


\section{Introduction}
\label{sec:intro}

The conventional theory for the origin of disk galaxies in Cold Dark
Matter (CDM) cosmologies involves the dissipational collapse of gas
inside relaxed dark matter halos formed through hierarchical
clustering \citep{white1978a,blumenthal1984a}.  This theory endows forming
disk galaxies with kinematic properties inherited from the halos that
host them \citep{fall1980a}.  Early simulations of cosmological
structure formation \citep{barnes1987a} verified that during linear
growth, the angular momentum of halos grows according to perturbation
theory \citep{peebles1969a}, and by incorporating these results
subsequent work explained the flat rotation curves and sizes of disk
galaxies through the combination of gas dissipation, halo spin, and
the adiabatic response of dark matter to baryonic infall
\citep{blumenthal1986a,mo1998a}.  The initial success of the dissipational
collapse model led to its establishment as the favored theory for the
formation of disk galaxies.  However, numerical modeling revealed that
the theory was incomplete.

While smooth, dissipational collapse may be relevant to disk galaxy
formation in quiet environments, simulations and semi-analytic modeling
have
demonstrated that it is rare for halos to acquire most of their angular
momentum via the quiescent accretion of tidally torqued material
\citep{vitvitska2002a,maller2002a,donghia2004a}.  
Simulation and observational work has shown that the
specific angular momentum content of dark matter halos differs from
exponential disks \citep{bullock2001a, van_den_bosch2001a, van_den_bosch2002a}
as halos contain comparatively more low angular momentum material that might produce
bulges or overly centrally concentrated disks \citep{van_den_bosch1998a,
van_den_bosch2001b}.
Furthermore, simulations of mass
accretion onto galaxies indicate that the hierarchical nature of the
$\Lambda$CDM cosmology leads to the destruction of disks when
dissipative effects are neglected \citep{toth1992a,quinn1993a,walker1996a,velazquez1999a}.
Models of galaxy collisions including dissipation
\citep{hernquist1989a,barnes1991a,barnes1996a} show that gas can loose
angular momentum owing to gravitational torques during
these events.  In cases when the
interstellar medium (ISM) is isothermal and relatively cold and the
gas fraction of the galaxies is small ($\sim 10\%$ of the baryons),
the subsequent inflow of gas into the centers of the merger remnants
forms a roughly spherical stellar distribution through a luminous
starburst \citep{mihos1994c,mihos1996a}, leaving objects that have
essentially no extended stellar disks.  Cosmological simulations of disk
formation with similar physics produce galaxies that are too centrally
concentrated, form too many stars, contain overly dominant bulges, and
lack angular momentum compared with nearby spirals
\citep{katz1991a,katz1992a,navarro1994a,steinmetz1995a,navarro2000a}.  

Efforts
to resolve these discrepancies in a cosmological context have mainly
relied on increasing the impact of feedback from star formation
\citep{weil1998a,sommer-larsen1999a,sommer-larsen2003a,thacker2000a,thacker2001a,abadi2003a,governato2004a,robertson2004a,okamoto2005a}.
Cosmological simulations that utilize strong feedback
from star formation in the form of a pressurized ISM to produce
disk galaxies either produce too few exponential disks
compared with observations \citep{robertson2004a} or yield
galaxies that still contain
a significant bulge component
\citep{abadi2003a}.
While these recent simulations have had some success, taken together the
results suggest that
the theory of structure formation remains faced with an apparent
incompatibility between models of disk formation and the hierarchical
growth of structure.

To help resolve this incompatibility, we propose a new, merger-driven
theory for the formation of disk galaxies at high redshifts that 
supplements standard dissipational collapse.
In our scenario, gas-dominated mergers, where the interacting systems
have a gas fraction $\fgas \equiv \Mgas/(\Mgas+\Mstar) \gtrsim 0.5$
at the height of the merger, allow for the formation of
rotationally-supported disks in remnants if energetic feedback
mechanisms limit the conversion of gas into stars.  
Gas pressurization from a multiphase ISM by
feedback from star formation \citep{mckee1977a,springel2003a} stabilizes
gas disks and allows mergers between gas-dominated galaxies to produce 
large,
smoothly distributed, rapidly rotating stellar disks.
Our proposed scenario builds upon the results of previous simulations 
of remnant disk 
formation in mergers by \cite{barnes2002a} and \cite{springel2005a}.

The merger-driven scenario for disk formation transitions naturally
into the merger-driven scenario for elliptical galaxy formation 
\citep[e.g.][]{toomre1977a} as star formation in galaxies over time
reduces the
gas content of progenitors and prevents the further occurrence of
mergers involving gas-dominated systems.  Mergers between gas-rich
disk galaxies at intermediate redshifts ($z\approx1-3$) are still
expected to produce realistic elliptical galaxies nearly universally
\citep[see e.g.][]{robertson2006a}, as prior quiescent star formation
and interactions will decrease their gas content below the necessary
$\fgas \gtrsim 0.5$ needed to produce a disk-dominated remnant.

Below we describe the simulation methodology used to 
demonstrate the merger-driven formation of disk galaxies 
(\S \ref{section:methodology}), including our methods
for studying the effects of
feedback (\S \ref{subsection:methodology:feedback}), orbital
geometry (\S \ref{subsection:methodology:orbits}), gas
fraction (\S \ref{subsection:methodology:gas_fraction}), merger
mass-ratio (\S \ref{subsection:methodology:minor_mergers}), and
progenitor mass (\S \ref{subsection:methodology:small_mergers})
on remnant properties.  Our method for analyzing the structural
and kinematic properties of the remnants is detailed in 
\S \ref{subsection:methodology:analysis} and the results of the
numerical experiments are presented in \S \ref{section:results}.
We discuss and summarize our results in \S \ref{section:discussion}
and \S \ref{section:summary}.

\section{Methodology}
\label{section:methodology}

To demonstrate the feasibility of our scenario
for merger-driven disk formation,
we have performed a suite of merger simulations of isolated, gas-rich
disk galaxies, with and without feedback from accretion onto
supermassive black holes (BHs).  The simulations use Smoothed Particle Hydrodynamics
(SPH) \citep{lucy1977a,gingold1977a} to evolve the gas and incorporate a
multiphase ISM model
\citep{yepes1997a,hultman1999a,springel2003a,springel2005b} where we control
the pressurization $\qeos$ of star-forming gas by interpolating
between isothermal gas ($\qeos=0$) with a temperature $T_{\rm
eff} = 10^{4}$ K and a stiffer multiphase medium ($\qeos=1$)
with a higher effective temperature $T_{\rm eff} \approx 10^{5}$ K by
varying the equation of state \citep[EOS; for details see][]{springel2005b}.  
Star formation in the dense ISM
($\Sigma > 10 \Msun \mathrm{\,pc}^{-2}$) occurs on a timescale
chosen to match observations \citep{kennicutt1998a}.  BH growth is
modeled \citep{springel2005b} by spherical accretion onto BH ``sink''
particles, with a fraction $\epsilon_{\rm f}$ of the released
gravitational potential energy of the accreted gas being deposited as
thermal feedback into the gas surrounding the BH.  For the simulations
and accretion model considered here, $\epsilon_{\rm f} = \eta_{\rm
therm}\epsilon_{\rm r} = 0.005$, where $\epsilon_{\rm r} \approx 0.1$
is the fraction of rest mass energy radiated during the growth of
the BH (fixed by standard accretion theory)
and the thermal coupling efficiency $\eta_{\rm therm} = 0.05$
is chosen to reproduce the local $\MBH - \sigma$ relation
\citep{di_matteo2005a}.

We consider two merger progenitors, a $\Vvir=160$ km s$^{-1}$ system
\citep{springel2005a} 
with a rotation curve and virial
mass $\Mvir = 9.5 \times 10^{11} h^{-1} \Msun$ similar to
the Milky Way and a smaller, $\Vvir=80$ km s$^{-1}$ system with
virial mass $\Mvir = 1.2 \times 10^{11} h^{-1} \Msun$.
The galaxies are comprised of exponential gas and stellar disks
embedded in dark matter halos with a
\cite{hernquist1990a} profile having scale lengths corresponding to a
Navarro-Frenk-White \citep{navarro1997a} concentration of $c_{\rm
NFW}=9$.  The progenitors each contain $\Ngas=40000$ gas, 
$\Nstars=40000$ stellar, and
$\Ndm=180000$ dark matter halo particles, and may also include a BH seed of
$10^{5} h^{-1}\Msun$
that is allowed to accrete in the manner described above.  Once the
BH particles fall within our force resolution limit, the sink
particles combine into a single supermassive BH.

To characterize the conditions under which our proposed scenario 
for merger-driven disk formation is most successful, we simulate 
a variety of merger
scenarios varying the ISM pressurization ($\qeos$), gas fraction 
($\fgas$), merger mass-ratio, disk orientations, and orbits.
The complete suite 
of simulations is described below, and a summary is provided
in Table \ref{table:models}.  For convenience we will refer to
each model with a two letter designation, with the first letter
signifying the gas fraction $\fgas$ and the second letter 
noting the ISM pressurization $\qeos$.  When necessary, we
add suffixes to describe e.g. orbital or merger mass-ratio
variations.  The simulation designations are detailed in the following
sections.

\begin{deluxetable*}{lcccccccccccc}
\tablecaption{\label{table:models}Merger Models}
\tabletypesize{\scriptsize}
\tablecolumns{13}
\tablehead{Model & $\VvirA$ & $\VvirB$ & $\fgas$ & $\qeos$ & $\rperi$ & $\rinit$ & $\theta_{1}$ & $\phi_{1}$ & $\theta_{2}$ & $\phi_{2}$ & BH \\
& [km s$^{-1}$] & [km s$^{-1}$]  & & & [$h^{-1}$ kpc] & [$h^{-1}$ kpc] & [degrees] & [degrees] & [degrees] & [degrees] & }
\startdata
GA   & 160 & 160 & 0.99 & 0.1  &  6.0 &  70.0 &    0 &   0 &   0 &   0 & Y  \\
GB   & 160 & 160 & 0.99 & 0.25 &  6.0 &  70.0 &    0 &   0 &   0 &   0 & Y  \\
GC   & 160 & 160 & 0.99 & 0.5  &  6.0 &  70.0 &    0 &   0 &   0 &   0 & Y  \\
GD   & 160 & 160 & 0.99 & 0.75 &  6.0 &  70.0 &    0 &   0 &   0 &   0 & Y  \\
GE   & 160 & 160 & 0.99 & 1.0  &  6.0 &  70.0 &    0 &   0 &   0 &   0 & Y  \\
\cutinhead{No BHs}
GAn  & 160 & 160 & 0.99 & 0.1  &  6.0 &  70.0 &    0 &   0 &   0 &   0 & N  \\
GBn  & 160 & 160 & 0.99 & 0.25 &  6.0 &  70.0 &    0 &   0 &   0 &   0 & N  \\
GCn  & 160 & 160 & 0.99 & 0.5  &  6.0 &  70.0 &    0 &   0 &   0 &   0 & N  \\
GDn  & 160 & 160 & 0.99 & 0.75 &  6.0 &  70.0 &    0 &   0 &   0 &   0 & N  \\
GEn  & 160 & 160 & 0.99 & 1.0  &  6.0 &  70.0 &    0 &   0 &   0 &   0 & N  \\
\cutinhead{$\rperi$ Variations}
GCrA & 160 & 160 & 0.99 & 0.5  &  2.7 &  70.0 &    0 &   0 &   0 &   0 & Y  \\
GCrB & 160 & 160 & 0.99 & 0.5  & 16.0 &  70.0 &    0 &   0 &   0 &   0 & Y  \\
GCrC & 160 & 160 & 0.99 & 0.5  & 32.0 &  70.0 &    0 &   0 &   0 &   0 & Y  \\
GCrD & 160 & 160 & 0.99 & 0.5  & 48.0 & 140.0 &    0 &   0 &   0 &   0 & Y  \\
\cutinhead{Orbital Variations}
GCoC & 160 & 160 & 0.99 & 0.5  &  6.0 &  70.0 &  180 &   0 & 180 &   0 & Y  \\
GCoE & 160 & 160 & 0.99 & 0.5  &  6.0 &  70.0 &   30 &  60 & -30 &  45 & Y  \\
GCoF & 160 & 160 & 0.99 & 0.5  &  6.0 &  70.0 &   60 &  60 & 150 &   0 & Y  \\
GCoO & 160 & 160 & 0.99 & 0.5  &  6.0 &  70.0 & -109 &  30 &  71 & -30 & Y  \\
GCoP & 160 & 160 & 0.99 & 0.5  &  6.0 &  70.0 & -109 &  90 & 180 &   0 & Y  \\
\cutinhead{$\fgas$ Variations}
DC   & 160 & 160 & 0.4  & 0.5  &  6.0 &  70.0 &    0 &   0 &   0 &   0 & Y  \\
EC   & 160 & 160 & 0.6  & 0.5  &  6.0 &  70.0 &    0 &   0 &   0 &   0 & Y  \\
FC   & 160 & 160 & 0.8  & 0.5  &  6.0 &  70.0 &    0 &   0 &   0 &   0 & Y  \\
\cutinhead{Minor Mergers}
DCm  & 160 &  80 & 0.4  & 0.5  &  6.0 &  70.0 &    0 &   0 &   0 &   0 & Y  \\
ECm  & 160 &  80 & 0.6  & 0.5  &  6.0 &  70.0 &    0 &   0 &   0 &   0 & Y  \\
FCm  & 160 &  80 & 0.8  & 0.5  &  6.0 &  70.0 &    0 &   0 &   0 &   0 & Y  \\
\cutinhead{$\Vvir=80$ km s$^{-1}$ Progenitors}
DCs  &  80 &  80 & 0.4  & 0.5  &  6.0 &  70.0 &    0 &   0 &   0 &   0 & Y  \\
ECs  &  80 &  80 & 0.6  & 0.5  &  6.0 &  70.0 &    0 &   0 &   0 &   0 & Y  \\
FCs  &  80 &  80 & 0.8  & 0.5  &  6.0 &  70.0 &    0 &   0 &   0 &   0 & Y  \\
\enddata
\end{deluxetable*}

\subsection{Feedback Study}
\label{subsection:methodology:feedback}

To study the impact of energetic feedback mechanisms
on the formation of disks in gas-rich mergers, we
perform 12 simulations of major mergers between
$\Vvir=160$ km s$^{-1}$ galaxies with disks consisting
almost entirely of gas ($\fgas=0.99$, ``G" runs).  The ISM 
pressurization
is incrementally increased over the values $\qeos=0.1$, $0.25$, 
$0.5$, $0.75$, and $1.0$, stiffening the equation of state of 
the ISM and increasing the sound speed of the gas at a given
density and temperature.
The simulations are referred to as models GA-GE in Table 
\ref{table:models}.
  We perform each simulation twice,
once with and once without the inclusion of growing supermassive
BHs in the progenitor systems.
The models without BHs are listed with the suffix ``n" in
Table \ref{table:models}, and the model GCn ($\fgas=0.99$, $\qeos=0.5$)
corresponds to the
simulation performed by \cite{springel2005a}.

The galaxies merge
on coplanar, prograde orbits with an initial separation
$\rinit = 70 h^{-1}$ kpc and pericentric distance $\rperi
= 6 h^{-1}$ kpc.  The orbit for the feedback study runs was
chosen to reproduce the encounter simulated by 
\cite{springel2005a}.
By using the same orbit for these simulations,
we explore the importance of feedback for the formation
of disks in mergers separate from the role of angular momentum.
We explore the role of angular momentum in producing remnant disks
in mergers through additional simulations
described in \S \ref{subsection:methodology:orbits}.

\subsection{Orbital Study}
\label{subsection:methodology:orbits}

The original orbital and disk angular momentum of 
the progenitors will affect the ability
of disks to form from mergers as 
angular momentum present in residual gas left over from
the merger can lead to rapidly rotating structures
in the remnant.
The formation of
remnant stellar disks have been demonstrated 
in interactions with prograde, coplanar orbits 
\citep{springel2005a}, and in simulations without star formation
gaseous remnant disks have been shown to form in mergers
with both polar and inclined orbits \citep{barnes2002a}.

To judge the importance of orbital and disk
angular momentum on a merger-driven scenario for
disk galaxy formation, we perform a study of 9
additional simulations.  In four models, we 
vary the pericentric passage distance $\rperi$
of the merger to alter the orbital angular momentum
of the interaction.  Typically, increasing $\rperi$
will increase the angular momentum of the orbit and
lead to larger, more rotationally-supported disks.
However, mergers with larger $\rperi$ 
undergo inefficient dynamical friction, avoid 
strong angular momentum loss 
during the first passage, and therefore take longer to merge.
Quiescent star formation will reduce the effective
gas fraction of the progenitors before the disks
merge in these cases, and may lead to larger
stellar spheroids.  These two competing effects
will be present in cosmological environs to some
degree, and for this reason we do not attempt to 
adjust for the decrease in gas fraction in wide-orbit
mergers.   In addition to a coplanar orbit with
$\rperi = 6 h^{-1}$ kpc, which is 
approximately twice the progenitor disk scale length $\Rd$,
we simulate one model with a smaller pericentric 
passage distance ($\rperi=\Rd$, model GCrA in Table
\ref{table:models}) and three
with larger pericentric passage distances ($\rperi= 0.1 \Rvir$,
$0.2\Rvir$, and $0.4\Rvir$, models GCrB-GCrD in Table
 \ref{table:models}).

In the remaining five models of the orbital study,
we vary the orientation of the progenitor disks.
While varying the pericentric passage distance
increases the orbital angular momentum, much of 
this angular momentum must be lost to the dark
matter halo through dynamical friction before
the disks can merge.  The original disk angular 
momenta of the progenitors may then substantially
add to the angular momentum of the remnant disk.
In addition to the coplanar orbit considered by
\cite{springel2005a},
we select five additional
disk orientations from the \cite{barnes1992a} study
of equal-mass mergers of stellar disk galaxies.
These orientations range from retrograde-retrograde
coplanar ($\theta_{1}=180$, $\phi_{1}=0$,
$\theta_{2}=180$, $\phi_{2}=0$, model GCoC)
to prograde-retrograde polar
($\theta_{1}=60$, $\phi_{1}=60$, $\theta_{2}=150$, 
$\phi_{2}=0$, model GCoF), with a complete list
provided in Table \ref{table:models}.
The results of these simulations are presented in 
\S \ref{subsection:results:orbits}

\subsection{Gas Fraction Study}
\label{subsection:methodology:gas_fraction}

The production of remnant disks in mergers will depend on 
the gas fraction $\fgas$ of the progenitor disk systems.
Stellar disks present before the merger will be substantially
heated during the merger and form stellar spheroids.  Only 
galaxies with significant gas fractions before the merger will
be able to re-form gaseous disks after the galaxies collide.
To characterize the progenitor gas fraction necessary to 
form a substantial remnant disk, we perform three simulations
with growing BHs, ISM pressurization $\qeos=0.5$ and gas fractions 
$\fgas = 0.4$, $0.6$, and $0.8$ (models DC, EC, and FC in
Table \ref{table:models}).
The results of these simulations are presented in 
\S \ref{subsection:results:gas_fraction}

\begin{figure*}
\figurenum{1}
\epsscale{1}
\plotone{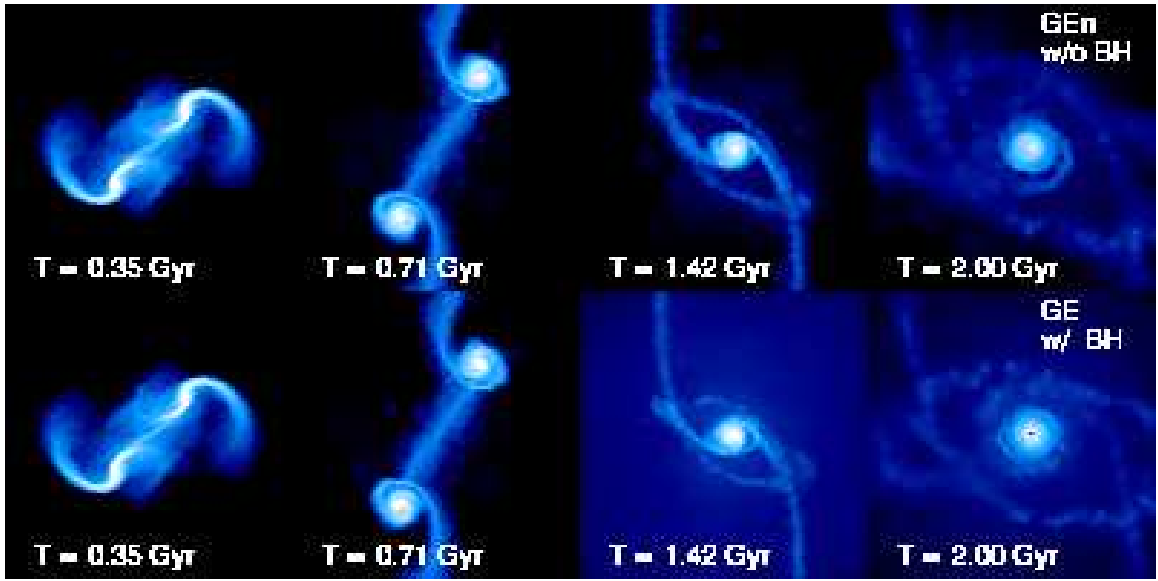}
\caption{\label{fig:time_sequence} 
Merger of two gas-rich disk galaxies, with (bottom, model GE) and without
(top, model GEn) supermassive black hole accretion and feedback.  Shown is the
gas surface density of the galaxies in each $140 \times 140$ kpc
square panel, demonstrating that the high-angular momentum merger
results in a remnant disk.  The effects of black hole feedback can be
seen over the 2 Gyr timescale of the merger as a diffuse, hot halo
and the lowered central gas density in the remnant (bottom panel).
}
\end{figure*}

\subsection{Minor Merger Study}
\label{subsection:methodology:minor_mergers}

In $\cdm$ cosmologies, the frequency of galaxy
mergers typically increases with the mass ratio of
the merging pair \citep[see e.g.][]{lacey1993a}.
As mentioned above, previous theoretical studies 
have suggested that dissipationless interactions
between satellites and disk galaxies can destroy
disks even if the satellite mass is relatively
small \citep[$\sim$ a few percent of the disk 
mass, e.g.][]{quinn1993a}.
Minor mergers with progenitor mass ratios
 of $3:1$ have been shown to produce small gaseous disks
in simulations without star formation \citep{barnes2002a}.
In simulations including gas dissipation and star formation,
minor mergers have been shown to induce starbursts 
\citep[e.g][]{mihos1994a}.
Cosmological simulations suggest that some
minor mergers may also add mass to the thick
component of disk galaxy systems \citep{abadi2003b}.
Minor mergers may then greatly influence the 
formation and survival of disk galaxies even
though the mass accretion rate from minor mergers
is typically smaller than that from major mergers.
To address the relevance of minor mergers for remnant
disks, we rerun the gas fraction study simulations from
\S \ref{subsection:methodology:gas_fraction} substituting a
$\Vvir = 80$ km s$^{-1}$ galaxy for the second progenitor.
These mergers with mass ratios of $m_{2}/m_{1}\sim8:1$ will
help characterize the influence of minor mergers on remnant disk
properties and provide examples of models that 
approximate cosmologically frequent merger events.
The results of these simulations are presented in
\S \ref{subsection:results:minor_mergers}.

\subsection{Small Progenitor Study}
\label{subsection:methodology:small_mergers}

The scale dependent physics of gas dissipation and star formation
may influence the ability of remnant disks to form in mergers.
Gas cooling is typically more efficient in smaller mass halos as
the virial temperature of those systems approach the regime where
hydrogen and helium recombination and collisional ionization 
enables large cooling rates \cite[see e.g.][]{black1981a}.  Gas
densities in the disks of low mass systems are typically lower
than in more massive systems, leading to longer gas consumption
timescales.  Quiescent star formation in low mass systems during
the phases of the merger when the separation between disks is large
and gravitational torques are weak will be lower than for high mass
systems, which may correspond to a larger gas fraction in the progenitor
disks when the systems actually merge \citep{robertson2006a}.

To gauge the net impact of these effects on the formation of remnant
disks in mergers of different mass scales, we perform three additional
simulations of equal mass mergers between $\Vvir=80$ km s$^{-1}$ galaxies
wit gas fractions of $\fgas=0.4$, $0.6$,
and $0.8$ for the progenitor systems (models DCs-FCs in Table \ref{table:models}).
The results of these simulations are presented in \S \ref{subsection:results:small_mergers}

\subsection{Analysis}
\label{subsection:methodology:analysis}

Each simulation is evolved over the timescale needed for the progenitor
systems to merge,
corresponding to $T=3.0-10.3$ Gyr depending on the orbit.
To illustrate the structure of the merging system during the
collision, Figure \ref{fig:time_sequence} shows the time evolution 
of the gas in
mergers with a strongly pressurized ($\qeos=1.0$) ISM with
(model GE, bottom panel) and without (model GEn, top panel) feedback 
from accretion onto
supermassive BHs.  Each panel shows a $140$ kpc $\times$ $140$ kpc
square area around the center of mass of the interacting system.  At
early times in the merger ($T=0.35$ Gyr, far-left panels), tidal
interactions between the galaxies distort them, but as significant
accretion onto the BHs has yet to occur, the overall gas distribution
remains similar between the simulations.
As the gas in the simulation
with BHs (bottom panels) begins to experience thermal feedback from BH
accretion, a hot halo of diffuse, low-angular momentum gas expelled 
through a wind surrounds the interacting galaxies.
By the end of the simulations at $T=2.0$
Gyr (far-right panels), rotating gas disks remain, eventually forming
stellar disks.  The remnant containing a BH has cleared out some gas
from the inner regions of the disk, reducing the central stellar density
and producing a diffuse hot gas envelope.

\begin{figure*}
\figurenum{2}
\epsscale{0.8}
\plotone{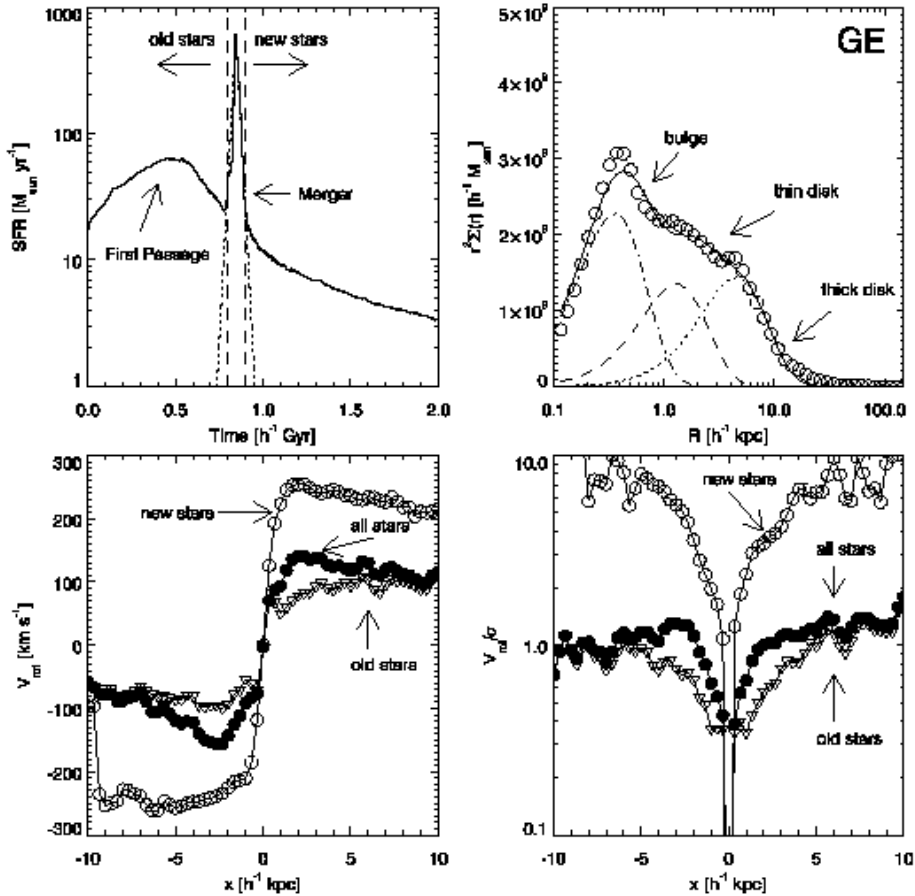}
\caption{\label{fig:analysis}
Example analysis performed for each merger simulation, 
shown for an equal mass merger between $\Vvir=160$ km s$^{-1}$ 
galaxies with gas-rich disks ($\fgas=0.99$), strongly 
pressurized ISM ($\qeos=1.0$), and growing supermassive 
black holes (model GE from Table \ref{table:models}).
The peak of the star formation rate (SFR) measured during
the simulation is used to identify the height of the merger
(upper left panel).  An exponential rise and decay is fitted to 
the SFR peak (dotted line) and used to categorize the stellar 
content of each remnant based on its formation time.  Stars
forming before three $e$-folding times prior to the
maximum of the SFR rate are labeled as ``old" stars 
(left dashed line), while
stars forming three $e$-folding times after the SFR peak
are labeled as ``new" stars (right dashed line).
Typically the new stars forming from remnant disks have
large rotational velocities (lower left panel) and are
rotationally supported with the ratio of rotationally velocity 
to velocity dispersion $\Vrot/\sigma > 1$ (lower right
panel).  The surface mass density $\Sigma(r)$ (upper right
panel) is modeled as a multi-component system consisting of
an exponential bulge, think disk, thick disk, and extended 
spheroid.  The model shown here has a surface mass density 
best modeled by a bulge--thin disk--thick disk system.
}
\end{figure*}

After the merger has completed, the structural and kinematic properties
of the remnant are measured from the simulation data.  The structural
components of the merger remnants, which might include a bulge, a thin
or thick disk, or an extended spheroid, are comprised of stars that
form at characteristically different times during the interaction.
The dynamically hot components of the remnant typically form before the
final coalescence of the merger, while the stars in the colder 
components of the remnant mostly form after the height of the merger
when the role of tidal forces has decreased.  The star formation rate (SFR)
of the merging system (see Figure \ref{fig:analysis}, upper left panel)
can be used define selection criteria for stars likely to populate 
separate dynamical components of the remnants based on their formation 
age.  The peak in the SFR that occurs during the height
of the merger can be approximated by an exponential rise and decay.
In what follows, we identify ``old'' stars in the remnants as stellar
particles forming earlier than before three $e$-folding times before the
center of the SFR peak and ``new'' stars as those forming later than
three $e$-folding times after the peak. 

The old and new stars in the remnant are then used to measure the
separate kinematics of the dynamically hot and cold stellar
structures of the remnant.  Typically, the old stars in
the remnant have relatively low rotational velocities 
(Figure \ref{fig:analysis}, lower left panel) and are not 
rotationally supported with the ratio of rotational
velocity to velocity dispersion $\Vrot/\sigma<1$ 
(Figure \ref{fig:analysis}, lower right panel).  New stars 
formed in the remnant after the height of the merger
typically have high rotational velocities and are 
rotationally supported.  Stars formed during the burst can
have kinematic properties intermediate between the old and
new stars, and therefore are not analyzed as
a separate stellar component.

\begin{figure*}
\figurenum{3}
\epsscale{1}
\plotone{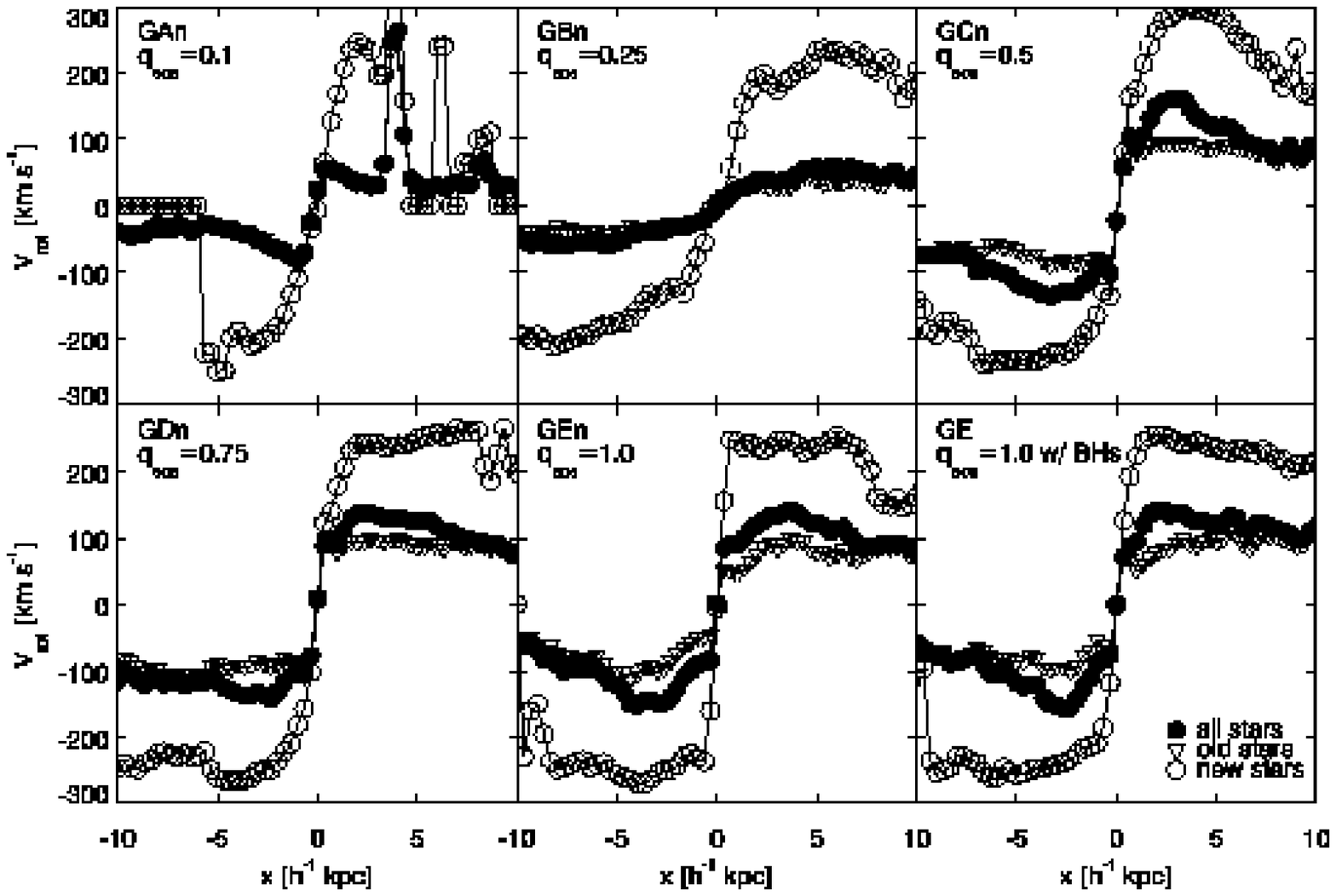}
\caption{\label{fig:v_feedback}
Stellar rotation curve of merger remnants, for models with
differing ISM pressurization.  Shown is the rotation of stars formed
before the merger (open triangles), after the merger (open circles),
and for all stars in the system (filled circles).  The amount of rotation
in the disk remnant correlates strongly with gas pressurization,
increasing from the upper left panel ($\qeos=0.1$, model GAn) to the bottom
middle panel ($\qeos=1.0$, model GEn).  The nearly isothermal ISM model
($\qeos=0.1$, model GAn) produces almost no net rotation in the stellar
disk, while moderately to strongly pressurized models ($\qeos
=0.5-1.0$, models GCn-GEn) yield a rotating stellar component.  The lower
right panel shows a strongly pressurized ISM model ($\qeos =
1.0$, model GE) with BHs.
}
\end{figure*}

The mass of the new stellar 
component forming in the remnant disk can be comparable or
large than the old, dynamically heated bulge or extended
spheroid.  The mass of the stellar structures in the remnant
can be measured from the surface mass density profile.
After the merger the stellar angular momentum of the remnant 
is measured and used to define a possible ``disk''-plane
perpendicular to the rotation axis of the remnant.  The plane
defined normal to the stellar angular momentum vector 
captures the mean stellar rotation in the remnants we study,
and only underestimates rotation when rotationally-supported
stellar structures are misaligned with the total stellar 
angular momentum.
The stellar surface mass density of each 
remnant is then measured in logarithmically-spaced annuli and 
is modeled by a multi-component system 
(Figure \ref{fig:analysis}, upper right panel) 
as a
function of radius $r$ that may include an exponential bulge 
\begin{equation}
\label{eqn:bulge}
\Sigma_{\mathrm{b}}(r) = \Sb e^{-r/\ab},
\end{equation}
\noindent
an exponential thin disk
\begin{equation}
\label{eqn:thin_disk}
\Sigma_{\mathrm{d}}(r) = \Sd e^{-r/\rd},
\end{equation}
\noindent
an exponential thick disk
\begin{equation}
\label{eqn:thick_disk}
\Sigma_{\mathrm{d,thick}}(r) = \Sdt e^{-r/\rdt},
\end{equation}
\noindent
and an extended stellar spheroid with a 
\cite{de_vaucouleurs1948a} profile
\begin{equation}
\label{eqn:spheroid}
\Sigma_{\mathrm{s}}(r) = \Ss 10^{-3.331\left[\left(r/\as\right)^{1/4} -1\right]}
\end{equation}
\noindent
Each
remnant is fitted with individual thin disk, bulge--thin disk, 
bulge--thin disk--thick disk, bulge--thin disk--spheroid, and 
bulge--thin disk--thick disk--spheroid models
with the requirements that the bulge scale length $\ab < 1$ kpc,
the thick disk scale length $\rdt$ is larger than the thin disk
scale length $\rd$, and the spheroid scale length $\as>\ab$.
A simple $\chi^{2}$ estimate of the success of each model in
reproducing the surface mass density profile of the remnant is
used to determine the relative mass in bulge, disk, and spheroid
components.  When the model fits suggest that
either a thick disk or an extended spheroid models provide 
comparable fits to the projected stellar distribution, the
rotational support $\Vrot/\sigma$ of the stars at large radii
is used to select a preferred model.  In this case, dispersion
supported systems ($\Vrot/\sigma<1$) are assigned extended
spheroids and rotationally-supported systems 
($\Vrot/\sigma\gtrsim1$) are assigned thick disk components.
The best-fit surface mass density models are reported in 
Table \ref{table:remnants}, including 
the remnant bulge-disk-spheroid mass ratios.  For
comparison, Table \ref{table:remnants}
also lists the best fit surface mass density parameters for 
a $\Vvir=160$ km s$^{-1}$ progenitor (model ICs) with a
gas fraction of $\fgas=0.4$ measured using the same method.

\section{Results}
\label{section:results}
The suite of simulations provides an extensive set of 
remnants that characterize the ability of extremely gas-rich
mergers to produce both disk components in spheroid-dominated
systems and disk-dominated systems whose structure and 
kinematics closely resemble large spiral galaxies.  In \emph{nearly every}
gas-rich merger we simulated (save model DCs, see below)
the new stars formed
after the peak of the SFR produced a rotationally-supported
structure.  These remnant disks are ubiquitous in gas-rich mergers,
but the extent to which they dominate the stellar content and the
degree of rotational support shows trends with the strength of
pressurization in the ISM, the presence of growing supermassive BHs,
progenitor gas fraction, orbit and disk orientation, and 
progenitor mass ratio.  Below, we consider the impact of
these merger properties in turn.

\begin{figure*}
\figurenum{4}
\epsscale{1}
\plotone{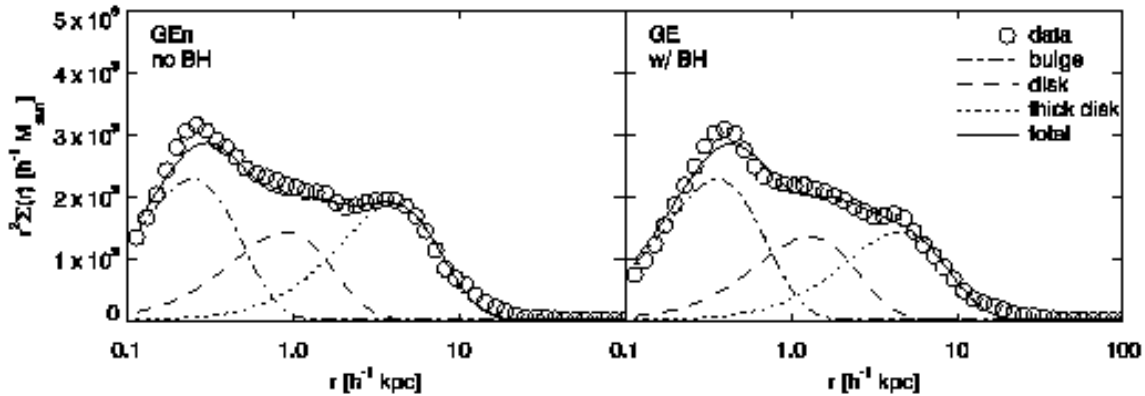}
\caption{\label{fig:smd_feedback}
Remnant stellar mass surface density with a strongly pressurized
($\qeos=1.5$) ISM, with (right panel, model GE) and without (left panel, 
model GEn) black hole feedback.  
 Shown is the measured stellar surface density (open circles), and
bulge (dot-dashed line), thin disk (dashed line), thick disk (dotted line),
and composite (solid line) mass models fits.
We fit exponential disks to the central bulge and stellar disk
components and a \cite{de_vaucouleurs1948a} profile to the spheroid.
As discussed by \cite{robertson2006a}, 
black hole feedback reduces slightly reduces the mass of the central
stellar spheroid and increases its effective radius.
 The relative sizes of the bulge and disk components are
reported in Table \ref{table:remnants}.
}
\end{figure*}

\begin{deluxetable*}{lccccccccccccc}
\tablecaption{\label{table:remnants} Merger Remnants}
\tabletypesize{\scriptsize}
\tablecolumns{14}
\tablewidth{0pc}
\tablehead{Model & $\Sb$ & $\ab$ & $\Sd$ & $\rd$ & $\Ss$ & $\as$ & $\Sdt$ & $\rdt$ & $\Mb$ & $\Md$ & $\Ms$ & $\Mdt$ & B:D:S}
\startdata
ICs & \nodata & \nodata & 4.9e8 & 2.73 & \nodata & \nodata & \nodata & \nodata & \nodata & 2.3e10 & \nodata & \nodata & 0.0:1.0:0.0\\
\cutinhead{BHs}
GA & 6.9e10 & 0.24 & 2.1e8 & 3.50 & 8.3e7 & 4.10 & \nodata & \nodata & 2.6e10 & 1.6e10 & 3.1e10 & \nodata & 1.6:1.0:1.9\\
GB & 3.9e10 & 0.41 & 2.1e8 & 4.48 & \nodata & \nodata & \nodata & \nodata & 4.2e10 & 2.6e10 & \nodata & \nodata & 1.5:1.0:0.0\\
GC & 2.0e11 & 0.17 & 1.6e9 & 1.55 & \nodata & \nodata & 3.3e7 & 5.19 & 3.8e10 & 2.5e10 & \nodata & 5.7e9 & 1.2:1.0:0.0 \\
GD & 7.3e10 & 0.18 & 1.9e9 & 1.42 & 8.0e7 & 2.89 & \nodata & \nodata & 1.6e10 & 2.4e10 & 1.5e10 & \nodata & 1.0:1.6:1.0 \\
GE & 1.3e11 & 0.18 & 5.9e9 & 0.65 & \nodata & \nodata & 5.2e8 & 2.26 & 2.7e10 & 1.6e10 & \nodata & 1.6e10 & 1.0:1.2:0.0\\
\cutinhead{No BHs}
GAn & 4.3e10 & 0.28 & \nodata & \nodata & 1.6e8 & 3.88 & \nodata & \nodata & 2.1e10 & \nodata & 5.6e10 & \nodata & 1.0:0.0:2.5\\
GBn & 5.0e10 & 0.28 & 8.2e7 & 4.45 & 3.7e8 & 1.97 & \nodata & \nodata & 2.6e10 & 1.0e10 & 3.3e10 & \nodata & 2.5:1.0:3.2\\
GCn & 1.8e11 & 0.18 & 1.3e9 & 1.61 & 1.8e7 & 4.15 & \nodata & \nodata & 4.1e10 & 2.1e10 & 7.3e9 & \nodata & 5.6:2.8:1.0\\
GDn & 1.9e11 & 0.18 & 1.2e9 & 1.48 & \nodata & \nodata & 1.4e8 & 3.21 & 4.1e10 & 1.6e10 & \nodata & 9.6e9 & 1.5:1.0:0.0\\
GEn & 2.7e11 & 0.12 & 1.3e10 & 0.45 & \nodata & \nodata & 8.5e8 & 2.02 & 2.8e10 & 1.7e10 & \nodata & 2.1e10 & 1.0:1.4:0.0\\
\cutinhead{Orbital Study}
GCoC & 7.0e10 & 0.25 & 1.5e9 & 1.78 & \nodata & \nodata & 3.0e7 & 5.39 & 2.7e10 & 3.0e10 & \nodata & 5.5e9 & 1.0:1.3:0.0\\
GCoE & 5.7e10 & 0.27 & 6.0e8 & 2.14 & 1.6e8 & 2.30 & \nodata & \nodata & 2.6e10 & 1.7e10 & 2.0e10 & \nodata & 1.5:1.0:1.1\\
GCoF & 4.7e10 & 0.25 & 1.7e9 & 1.96 & \nodata & \nodata & \nodata & \nodata & 1.9e10 & 4.2e10 & \nodata & \nodata & 1.0:2.1:0.0\\
GCoO & 7.2e10 & 0.22 & 1.5e9 & 1.94 & \nodata & \nodata & 9.3e6 & 5.59 & 2.3e10 & 3.7e10 & \nodata & 1.8e9 & 1.0:1.6:0.0\\
GCoP & 6.2e10 & 0.24 & 1.6e9 & 1.43 & 9.0e7 & 2.90 & \nodata & \nodata & 2.3e10 & 2.1e10 & 1.7e10 & \nodata & 1.3:1.2:1.0\\
GCrA & 5.7e10 & 0.32 & 2.0e9 & 1.28 & 2.8e7 & 4.29 & \nodata & \nodata & 3.7e10 & 2.1e10 & 1.1e10 & \nodata & 3.1:1.8:1.0\\
GCrB & 2.4e10 & 0.35 & 8.6e8 & 2.04 & \nodata & \nodata & 6.1e7 & 6.02 & 1.9e10 & 2.2e10 & \nodata & 1.3e10 & 1.0:1.8:0.0\\
GCrC & 4.0e9 & 0.54 & 1.5e8 & 3.32 & 1.0e8 & 4.31 & \nodata & \nodata & 7.5e9 & 1.0e10 & 4.4e10 & \nodata & 1.0:1.4:5.9\\
GCrD & \nodata & 0.67 & 4.6e7 & 5.57 & 2.0e8 & 3.42 & \nodata & \nodata & \nodata & 9.1e9 & 5.4e10 & \nodata & 0.0:1.0:5.9\\
\cutinhead{Gas Fraction Study}
DC & 3.3e10 & 0.25 & 2.7e7 & 6.15 & 1.6e8 & 3.65 & \nodata & \nodata & 1.3e10 & 6.6e9 & 4.9e10 & \nodata & 2.0:1.0:7.5\\
EC & 6.7e10 & 0.21 & 1.9e7 & 4.71 & 1.7e8 & 3.38 & \nodata & \nodata & 2.0e10 & 2.7e9 & 4.6e10 & \nodata & 7.6:1.0:17\\
FC & 2.4e11 & 0.11 & 7.7e9 & 0.54 & 2.9e7 & 4.89 & 2.7e8 & 2.77 & 1.9e10 & 1.4e10 & 1.6e10 & 1.3e10 & 1.1:1.7:1.0\\
\cutinhead{Minor Merger Study}
DCm & 4.3e9 & 0.57 & 3.7e8 & 3.36 & \nodata & \nodata & \nodata & \nodata &  9.4e9 & 2.6e10 & \nodata & \nodata & 1.0:2.8:0.0\\
ECm & 3.6e9 & 0.58 & 3.3e8 & 3.14 & 1.3e7 & 4.70 & \nodata & \nodata & 7.6e9 & 2.0e10 & 6.8e9 & \nodata & 1.1:3.0:1.0\\
FCm & 7.5e9 & 0.36 & 3.7e8 & 3.06 & 1.3e7 & 3.66 & \nodata & \nodata & 6.3e9 & 2.1e10 & 4.2e9 & \nodata & 1.5:5.1:1.0\\
\cutinhead{$\Vvir = 80$ km s$^{-1}$ Study}
DCs & 8.7e9 & 0.12 & 1.0e9 & 0.33 & 6.7e7 & 2.15 & \nodata & \nodata & 8.5e8 & 7.2e8 & 7.1e9 & \nodata & 1.1:1.0:9.8\\
ECs & 3.3e10 & 0.10 & 9.1e8 & 0.46 & 3.5e7 & 2.43 & \nodata & \nodata & 2.1e9 & 1.2e9 & 4.8e9 & \nodata & 1.7:1.0:3.9\\
FCs & 3.2e10 & 0.10 & 1.5e9 & 0.34 & 5.0e7 & 1.92 & \nodata & \nodata & 2.0e9 & 1.2e9 & 4.2e9 & \nodata & 1.7:1.0:3.4\\
\enddata
\end{deluxetable*}

\subsection{Feedback Study}
\label{subsection:results:feedback}

The kinematic properties of the remnants exhibit significant
quantitative differences as the pressure support of the ISM is varied.
The stellar rotation curves of the merger remnant in the $\qeos
=0.1,0.25,0.5,0.75,$ and $1.0$ models without BHs are plotted in
Figure \ref{fig:v_feedback} for a progenitor gas fraction of $\fgas=0.99$.
The strength of the rotating stellar component correlates with 
the pressurization of the ISM.  The least effective ISM model
for producing a remnant disk is the weakly pressurized,
nearly isothermal model ($\qeos=0.1$, model GAn, upper left panel)
that yields a slowly rotating, clumpy stellar remnant.
The gaseous progenitor disks in model GAn are not stable against
\cite{toomre1964a} instability as the sound speed of the weakly
pressurized gas cannot support the massive gas disk against
self-gravity.  The gaseous disk efficiently converts into
stellar clumps before and during the collision, producing a
remnant with very little average rotation 
($\max \Vrot \approx 90$ km s$^{-1}$).  However, the scant 
amount of gas remaining after the final coalescence in this
merger still manages to produce a very small, rapidly rotating 
disk.  Increasing the ISM pressurization to $\qeos=0.25$ 
(model GBn, upper middle panel) greatly reduces the presence
of stellar clumps owing to instabilities and roughly doubles the 
radial extent of the rotating component, but does little to 
improve the average rotation of the remnant.  Overall, 
low-pressurization ISM models do a poor job at producing
remnants with substantial average rotation but can produce
small, rapidly rotating remnant disk components.

\begin{figure*}
\figurenum{5}
\epsscale{1}
\plotone{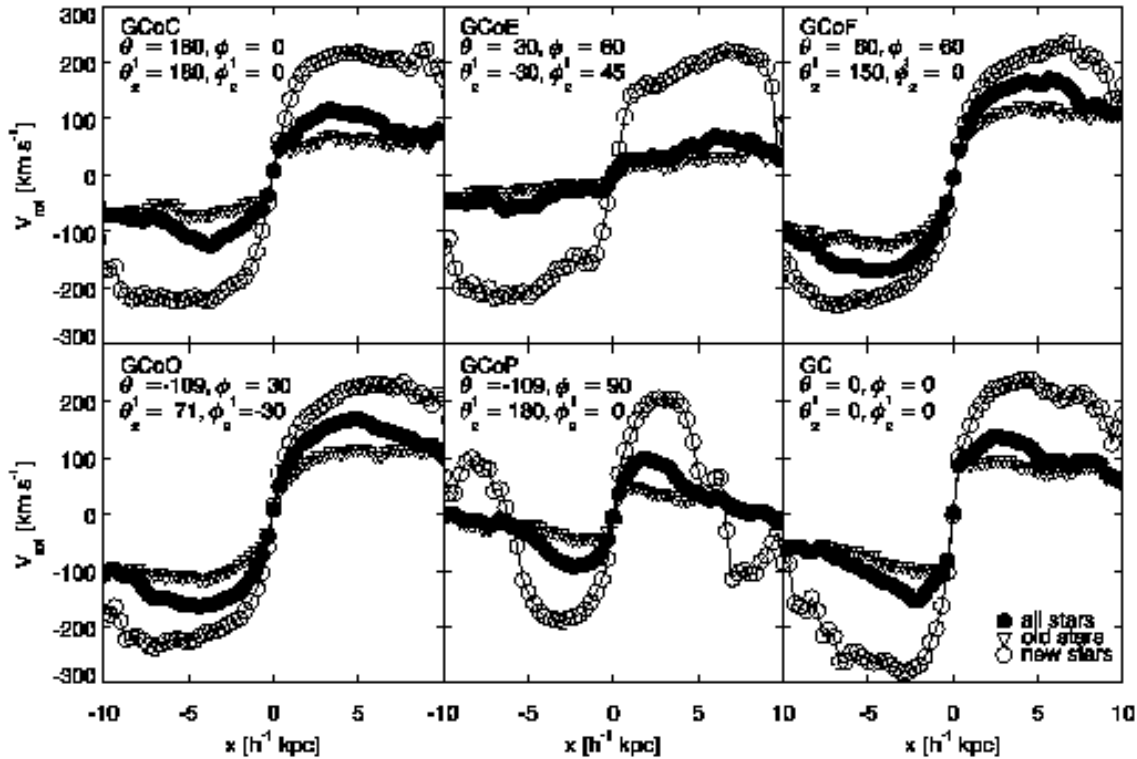}
\caption{\label{fig:v_orientations}
Rotation curves of remnants from mergers of galaxies with 
moderately pressurized ISM and growing supermassive black 
holes for a variety of disk orientations.  Shown is the
rotation for new stars formed after the merger (open circles),
old stars formed before the merger (open triangles), and
all stars (solid circles).  The formation of rapidly rotating
remnants is not limited to prograde-prograde coplanar mergers 
(model GC, lower right panel), but can occur in both polar 
(model GCoF, upper right panel) and inclined (model GCoO, lower
left panel) orbits.  Gas-rich mergers can also produce
remnants with very little rotation (model GCoE, upper middle
panel) or counter-rotating cores (model GCoP, lower middle panel).
}
\end{figure*}

Merger-driven disk formation will require a moderately to strongly
pressurized ISM to provide enough kinematic stability to the gas
to avoid forming too many stars before the final merger.
The moderately to strongly pressurized models 
($\qeos=0.5-1.0$, GCn-GEn, upper left-lower middle panels)
produce remnants that contain rapidly rotating stellar disks in
addition to bulge or extended spheroid components.  During the
merger, the pressurization from the stiffer equation of state
makes the ISM less compressible, reduces the central gas density, 
and leads to less conversion of gas into stars.  A larger fraction
of stars in the final remnant form after the height of the merger,
increasing the mass of the disk component and the average rotational
velocity of the stars.  Just before the final coalescence, the moderately 
and strongly pressurized models have $\fgas \approx 0.5-0.6$, much larger
than the $\fgas \approx 0.1$ in the $\qeos=0.1$ model and $\fgas=0.26$ in
the $\qeos=0.25$ model.
The abundance of gas present during the final coalescence in the moderately
and strongly pressurized models allows for stellar disks to form
from gaseous remnant disks \citep[e.g.][]{springel2005a}.
Figure \ref{fig:smd_feedback} shows the stellar surface mass density of two 
highly pressurized ($\qeos=1.0$), gas-rich major merger models with (model
GE, right panel) and without (model GEn, left panel) supermassive BH feedback.
In both cases, stellar disks form from gaseous remnant disks created
during the merger.   The presence of supermassive BHs slightly decreases the
mass of the central spheroid and increases its effective radius, in agreement
with the findings of \cite{robertson2006a} for mergers that produce elliptical 
galaxies.  Combined, the velocity fields
shown in Figure \ref{fig:v_feedback} and the stellar disks
shown in Figure \ref{fig:smd_feedback}
demonstrate that disk galaxy remnants
containing rapidly rotating, stellar disk components can form from
high-angular momentum, gas-rich mergers when the ISM is pressurized.

\subsection{Orbital Study}
\label{subsection:results:orbits}

While the results of \S \ref{subsection:results:feedback} demonstrated that
the pressurization of the multiphase ISM from star formation and feedback
can produce remnant disks in gas-rich, prograde-prograde coplanar mergers,
the generality of a merger-driven scenario for disk galaxy formation
would require a wider range of encounters to produce remnant disks.  
Using simulations without star formation,
\cite{barnes2002a} demonstrated that gaseous disks could form in mergers
that were not exactly coplanar, including polar orbits.
Here, we examine the impact of varying a range of 
progenitor disk orientations and pericentric passage distances on remnant 
disk formation including the effects of star formation and energetic
feedback mechanisms.

Figure \ref{fig:v_orientations} shows the rotation curves for six 
different fairly radial encounters, including a prograde-prograde 
coplanar merger (model GC, lower right panel),
retrograde-retrograde coplanar (model GCoC, upper left panel), polar 
(models GCoF, upper right panel, and GCoP, middle lower panel), and 
intermediate cases (models GCoE, upper middle panel, and GCoO, lower left
panel).  In each case, the same
moderately-pressurized ISM ($\qeos=0.5$) is used and enables the new 
stellar component to rapidly rotate.  Remarkably, some highly non-coplanar
encounters produce remnants whose entire stellar remnant is rapidly 
rotating (models GCoF and GCoO) and demonstrate that merger-driven disk
formation is not limited to the single coplanar orbit discussed by 
\cite{springel2005a}.  Certainly, some orbits lead to 
remnants with almost no average rotation (model GCoE), though with a 
rapidly rotating young stellar component.  Interestingly, the 
retrograde-retrograde polar encounter (model GCoP) produces a remnant with
a new stellar component that counter-rotates in its interior relative to 
the outer young stellar material.  Previous studies have demonstrated the
formation of counter-rotating disks to form in remnants of mergers 
involving progenitors with $\fgas\approx0.1$ \citep{hernquist1991a} or  
in spiral galaxies via other mechanisms 
\citep[e.g.][]{thakar1996a,thakar1998a}.
We plan to examine the 
creation of counter-rotating cores through extremely gas-rich major
mergers in subsequent work, but we note here that the relative ages of the
older, central disk in model GCoP compared with its outer, younger disk is
consistent with the picture for the formation of kinematically decoupled
cores presented by \cite{hernquist1991a} where the central disk forms 
primarily from
strongly-torqued material during the first passage and the outer disk 
forms later from material that retains some of its original orbital or
disk angular momentum.

\begin{figure*}
\figurenum{6}
\epsscale{1}
\plotone{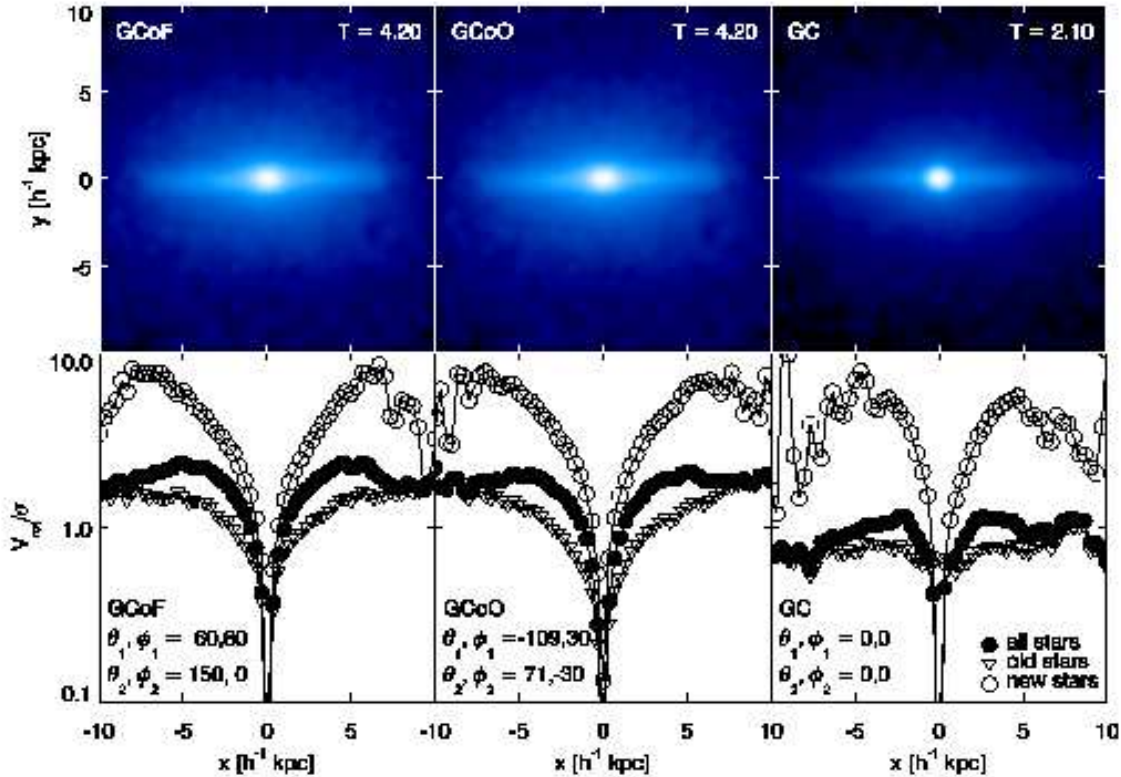}
\caption{\label{fig:vs_orientations}
Rotational support $\Vrot/\sigma$ and projected stellar mass
for three orbits that produce remnant disks in very gas-rich 
mergers.  Shown are the logarithmic surface density of the
stars (upper panels) and the $\Vrot/\sigma$ measured for 
all stars (solid circles), and stars formed before (open 
triangles) or after (open circles) the merger of galaxies with
moderately pressurized ISM models ($\qeos=0.5$).  The polar
(model GCoF, left panels) and inclined (model GCoO, middle
panels) orbits lead to remnants that are disk-dominated and
rotationally-supported.  The prograde-prograde coplanar
orbit (model GC, right panels) also produces a 
rotationally-supported disk system.
}
\end{figure*}

The rotational support of the remnants GCoF and GCoO formed from non-coplanar
mergers, shown in Figure 
\ref{fig:vs_orientations} with the coplanar encounter GC, is surprising but 
bolsters
the merger-driven scenario for disk formation.  In each merger, the total
stellar rotational support exceeds $\Vrot/\sigma = 1$ and the new stellar 
component peaks at $\Vrot/\sigma>5$.  While the mass of the 
prograde-prograde coplanar merger remnant is still dominated by the 
central bulge (bulge-to-disk ratio $B:D \approx 1.2:1.0$), the other 
models are $disk$-dominated with bulge-to-disk ratios of 
$B:D\approx1.0:2.1$ for model GCoF and $B:D\approx1.0:1.6$ for model 
GCoO including thick
disk components.  At the time of the merger these non-coplanar mergers 
have $\fgas\approx0.6-0.7$ of their baryonic component in gas, compared 
with $\fgas\approx0.5$ for the coplanar merger.  The 
rotationally-supported
remnants from non-coplanar orbits experience less star formation during 
their first passage than the coplanar orbit \citep[see also][]{mihos1996a},
leaving more gas available to
form a remnant disk after the final coalescence.  The inclined merger 
model GCoE (Figure \ref{fig:v_orientations}, upper middle panel) that 
produces
a remnant with very little net rotation has a $\fgas\approx0.2$ 
immediately after the final coalescence, whereas the disk remnants formed
by models GCoF and GCoO have $\fgas\approx0.5$ and $\fgas\approx0.4$, 
respectively.
Clearly, both the amount of gas present in the interacting system before 
the height of merger and the fraction of gas consumed during the final
coalescence may influence the formation of remnant disks by improving the
rotational support of the remnant.

The simulations that explore the effects of disk 
orientation demonstrate that for a given amount of orbital angular 
momentum in a merger the contribution of disk angular momentum can affect
the presence of a remnant disk, both by providing angular momentum to the
remnant gaseous disk that forms the rapidly-rotating stellar component and
by influencing the amount of gas consumed during the merger.  However, 
the orbital configurations considered above only alter the disk 
orientation and not the orbital angular momentum of the encounter.  
Changing the pericentric passage distance can influence both the angular
momentum of the remnant and the amount of gas available to form a remnant
disk both by changing the timescale of the merger and by affecting the 
gas consumption during the interaction. 

\begin{figure*}
\figurenum{7}
\epsscale{1}
\plotone{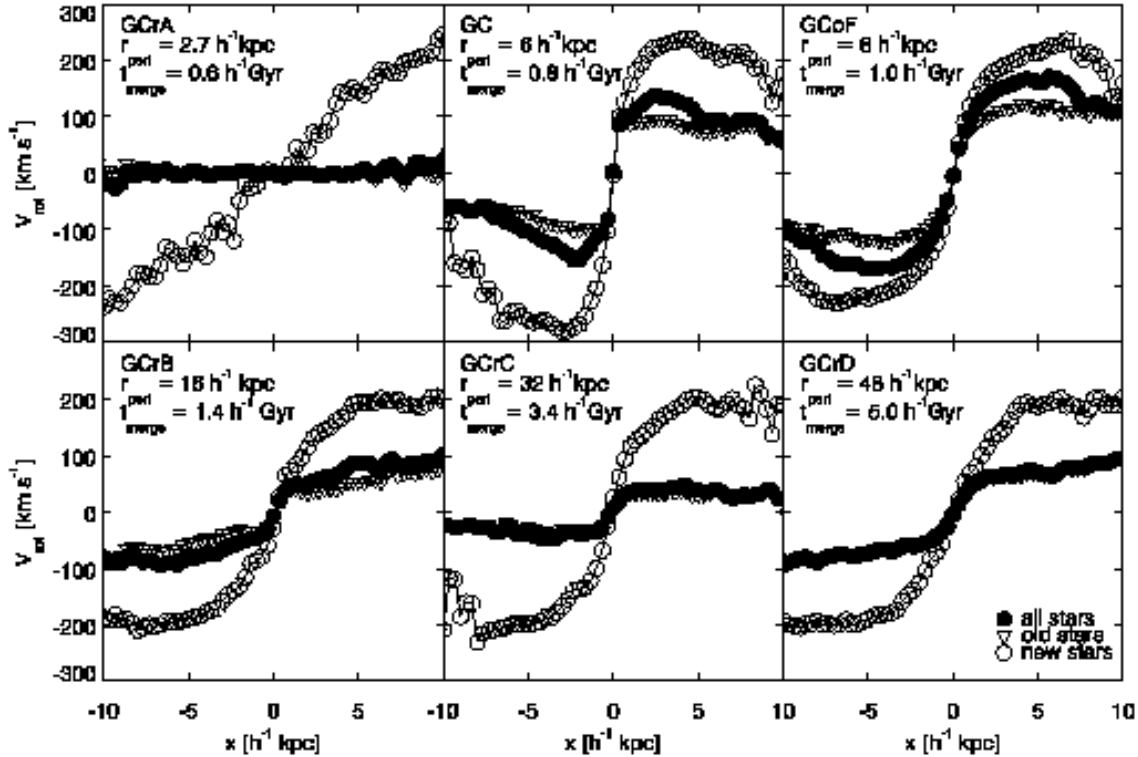}
\caption{\label{fig:v_rperi}
Rotation curves of remnants from mergers of galaxies with 
moderately pressurized ISM and growing supermassive black 
holes for a variety of pericentric passage distances.  Shown is the
rotation for new stars formed after the merger (open circles),
old stars formed before the merger (open triangles), and
all stars (solid circles) for pericentric passage distances
increasing from $\rperi =\rd=2.7h^{-1}$ kpc (model GCrA, upper left
panel) to $\rperi=0.3\Rvir=48 h^{-1}$ kpc (model GCrD, lower right
panel).  The formation of rapidly rotating
remnants is appears to require orbits that are not too radial, since
the first passage is violent and too much gas is converted to stars
before the merger.  Orbits also cannot be too wide, since the merger
timescale is extended and results in a mostly stellar final merger.
Remnants from mergers with pericentric passage distances of a few progenitor disk 
scale lengths, such as the coplanar (model GC, upper middle panel) and
polar orbits shown here (model GCoF, upper right panel), appear to be
most efficient at producing remnant disks.
}
\end{figure*}

Figure \ref{fig:v_rperi} shows the rotation curves of remnants with 
moderately pressurized ISM models ($\qeos=0.5$) formed
in prograde-prograde coplanar gas-rich ($\fgas=0.99$) mergers with 
pericentric passage distances 
increasing from 
$\rperi=R_{\mathrm{d,progenitor}} = 2.7 h^{-1}$ kpc (model GCrA, upper 
left panel) to $\rperi=0.3R_{\mathrm{vir,progenitor}}=48 h^{-1}$ kpc 
(model GCrD, lower right panel).  Also shown for comparison is a polar 
merger with $\rperi\approx 2R_{\mathrm{d,progenitor}}$ (upper right 
panel), the same pericentric passage distance as the prograde-prograde 
coplanar model GC (upper middle panel).  The nearly head-on 
collision (model GCrA) produces a remnant with almost no net rotation; 
after the height of the merger, only a small amount of gas remains 
($\fgas=0.13$).  The gas available for a remnant disk increases to
$\fgas\approx0.5-0.6$ when the pericentric passage disk scale length is 
roughly
twice the disk scale length (models GC and GCoF), and decreases towards 
larger $\rperi$ with $\fgas\approx0.2-0.4$ (models GCrB-GCrD).  These
orbits have a substantially longer merger timescale than the more radial
orbits, allowing for more quiescent star formation and a larger stellar
component present in the progenitors at the height of the merger.  The
competing effects of increasing the available orbital angular momentum
by increasing $\rperi$ and decreasing the gas available to form a remnant
disk owing to extended quiescent star formation then lead to a range of 
$\rperi$ from which remnant disks are likely to form.  While additional
simulations will be needed to better characterize the best orbits for
remnant disk formation, from the simulations performed here we estimate 
that fairly radial orbits with pericentric passage distances a few times 
larger than the disk scale length are most ideal for merger-driven
disk formation in gas-rich mergers.

\subsection{Gas Fraction Study}
\label{subsection:results:gas_fraction}

The preceding feedback and orbital studies have demonstrated that 
gas-rich mergers with a pressurized ISM can produce remnant disks for
a variety of orbital configurations.  The results of those simulations
suggest that the primary influence on the presence of remnant disks is
the amount of gaseous material left after the final coalescence; 
increased feedback or favorable (though diverse) orbits reduce the SFR
during the first pericentric passage and the height of the merger, leaving
enough gas after the merger to form a rapidly-rotating stellar component.
By altering directly the gas fraction of the progenitor systems, the role 
of gas fraction on the rotational support of merger remnants can be
determined.

\begin{figure*}
\figurenum{8}
\epsscale{1}
\plotone{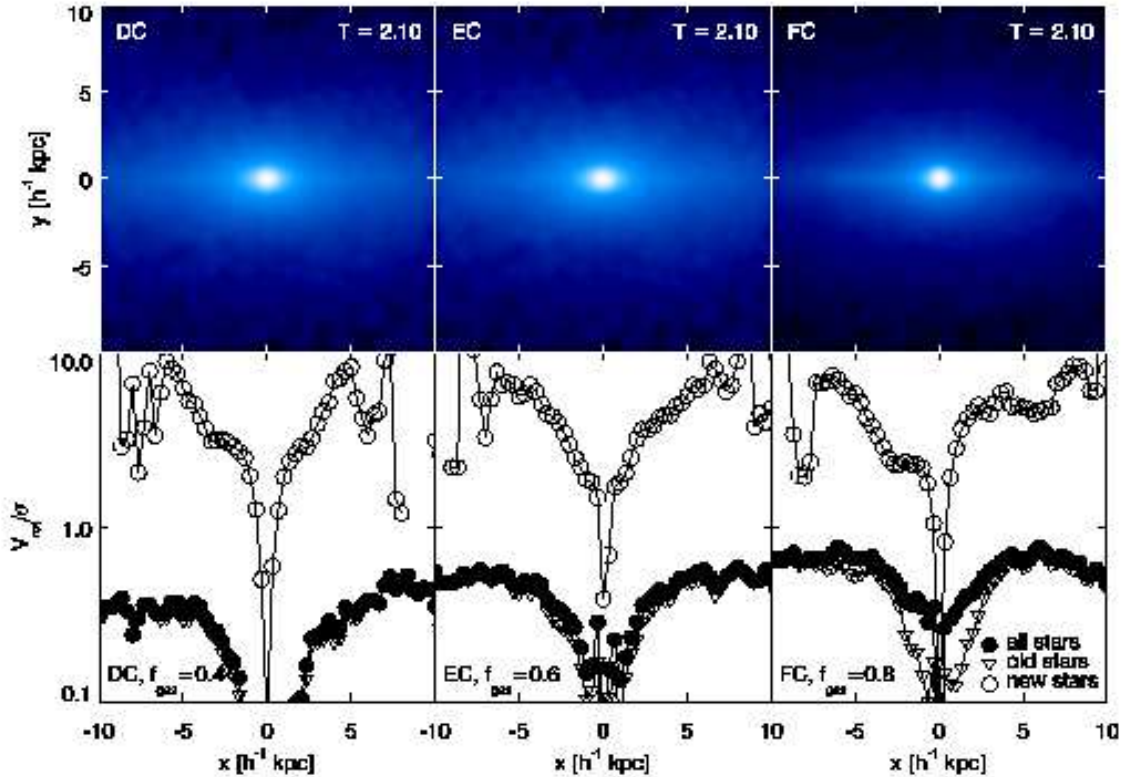}
\caption{\label{fig:vs_gas_fraction}
Rotational support $\Vrot/\sigma$ and projected stellar mass
for mergers with progenitor gas fractions ranging 
from $\fgas=0.4$ (model DC, left 
panel) to $\fgas=0.8$ (model FC, right panel).
mergers.  Shown are the logarithmic surface density of the
stars (upper panels) and the $\Vrot/\sigma$ measured for 
all stars (solid circles), and stars formed before (open 
triangles) or after (open circles) the merger of galaxies with
moderately pressurized ISM models ($\qeos=0.5$).  The rotational
component of these dispersion dominated galaxies correlate with
the gas fraction, and progenitor galaxies will likely need an initial
gas fraction of $\fgas>0.8$ to produce remnant disks (see 
Figure \ref{fig:v_feedback}).
}
\end{figure*}

Figure \ref{fig:vs_gas_fraction} shows the impact of increasing the gas
fraction from $\fgas=0.4$ (left panel, model DC) to $\fgas=0.8$ (right
panel, model FC) in the progenitor disks.  Shown are prograde-prograde 
coplanar mergers with a moderately pressurized ISM model ($\qeos=0.5$).
While the new stars in each
model produce a rotationally-supported structure, the average
rotational support of the system is well correlated with the gas fraction
as judged from the increasing $\Vrot/\sigma$ as a function of progenitor
$\fgas$.  The gas fraction of the merging galaxies before and after the
final coalescence are nearly in proportion to the original gas progenitor
gas fractions, with $\fgas=0.2$ before and $\fgas=0.1$ after the final
merger in model DC and roughly twice that those values for the FC model
with double the original gas content.  However, the original $\fgas=0.8$ 
progenitors in the model FC merger are not gas-rich enough to produce a
remnant that has a rotationally-supported total stellar component with
this orbital configuration.  These simulations suggest that given the
typical $\sim$ Gyr timescales involved, major merger progenitors must have
original gas fractions $\fgas \gtrsim 0.8$ in order to satisfy the
$\fgas\gtrsim0.5$ requirement during final coalescence for a 
disk-dominated remnant.

\subsection{Minor Merger Study}
\label{subsection:results:minor_mergers}

The simulations of 
\S \ref{subsection:results:feedback}-\ref{subsection:results:gas_fraction}
have demonstrated that gas-rich mergers can lead to remnant disk systems
with substantial rotational support.  However, these simulations have
modeled only equal mass encounters.  While major mergers on average
contribute significantly to the final mass of an average galaxy at the
present day, minor mergers between unequal mass galaxies are 
cosmologically more frequent \citep{lacey1993a}.  

The survival of disks in minor merger or infall scenarios have been 
readily 
addressed both semi-analytically \citep{toth1992a,benson2004a} and 
numerically 
\citep{quinn1993a,walker1996a,velazquez1999a,font2001a,barnes2002a}.  
Recently, 
\cite{bournaud2005a} has simulated comparatively gas-poor ($\fgas<0.2$) 
mergers between 
galaxies with mass ratios in the range $4:1-10:1$ which produce remnants 
with
disk-like stellar structures but elliptical-like kinematics 
\citep[see also][]{bournaud2004a}.  Currently, the combined results of 
these analyses suggest that stellar minor mergers with disks will
produce remnant systems that are kinematically dispersion-dominated but 
the cosmological infall of satellites expected in the $\lcdm$ cosmology is
consistent with the observed distribution of thicknesses for disk 
galaxies.

\begin{figure*}
\figurenum{9}
\epsscale{1}
\plotone{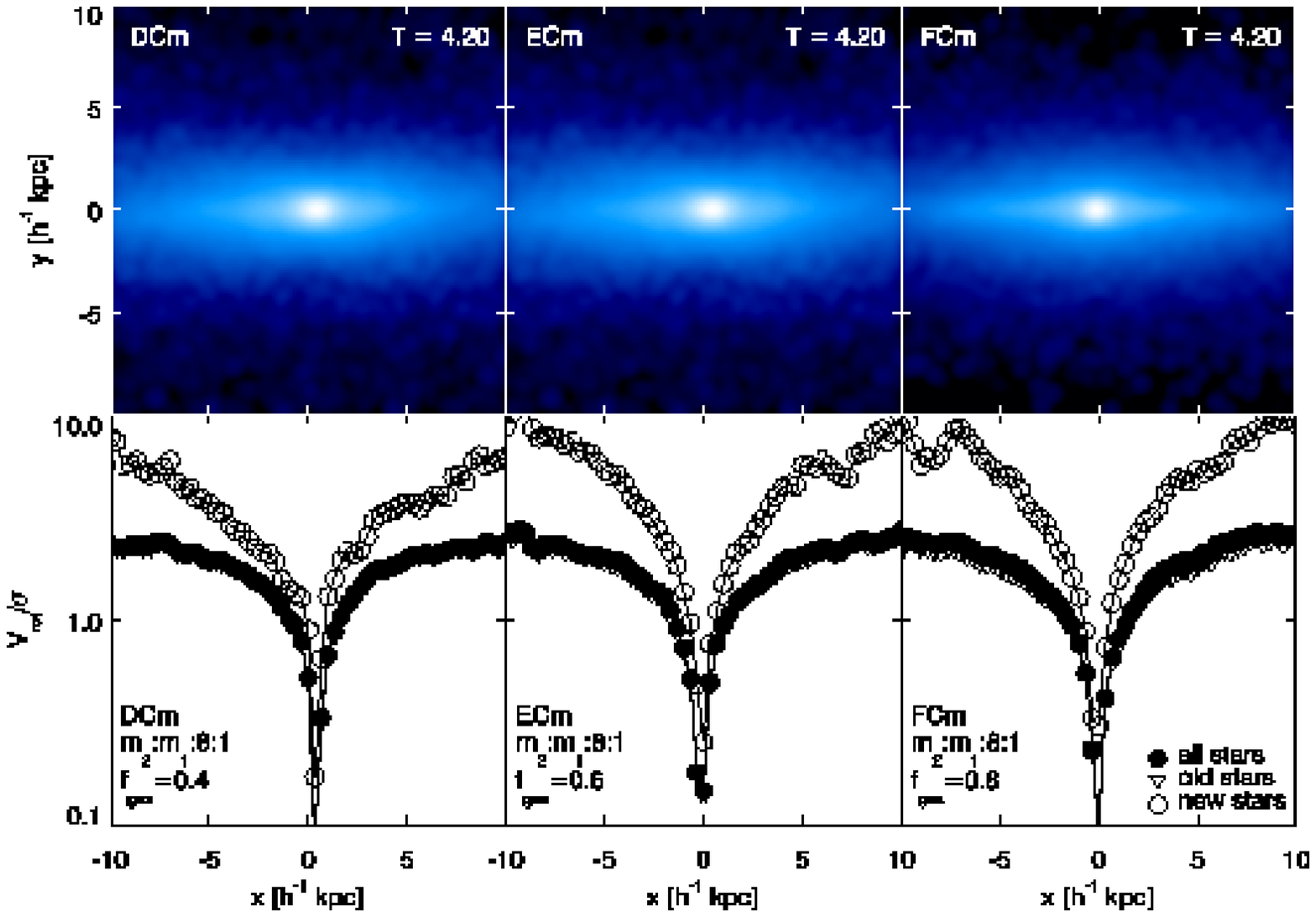}
\caption{\label{fig:vs_minor}
Rotational support $\Vrot/\sigma$ and projected stellar mass
for three minor mergers (mass ratio $m_{2}:m_{1}=8:1$)
for progenitor gas fractions ranging from
$\fgas=0.4$ (model DCm, left panel) to $\fgas=0.8$ (model FCm,
right panel).
Shown are the logarithmic surface density of the
stars (upper panels) and the $\Vrot/\sigma$ measured for 
all stars (solid circles), and stars formed before (open 
triangles) or after (open circles) the merger of galaxies with
moderately pressurized ISM models ($\qeos=0.5$).  The minor
mergers produce disk-dominated, rapidly-rotating remnants with
$\Vrot/\sigma>2$.  The central velocity dispersion of the larger
galaxy roughly double during the merger, but the rotational support
of the remnants are considerably larger than the $\Vrot/\sigma$
measured for comparable gas-poor minor mergers \citep[e.g.][]{bournaud2005a}. 
}
\end{figure*}

Of additional interest to a scenario for merger-driven disk 
formation, \cite{barnes2002a} demonstrated in hydrodynamical simulations 
without star formation that gas-rich minor mergers can produce small 
gaseous disk remnants.   A cosmological scenario for merger-driven disk 
formation would likely involve such minor mergers given their 
abundance in the mass-accretion histories of galaxies.  Figure 
\ref{fig:vs_minor} shows the rotational support of remnants formed in
prograde-prograde coplanar mergers of the 
$\Vvir=160$ km s$^{-1}$ galaxy model considered in 
\S \ref{subsection:results:feedback}-\ref{subsection:results:gas_fraction}
with the $\Vvir=80$ km s$^{-1}$ galaxy described in 
\S \ref{subsection:methodology:minor_mergers}.  These
minor mergers with mass ratio $m_{2}:m_{1}\approx8:1$ use a moderately
pressurized ISM ($\qeos=0.5$) and progenitor gas fractions increasing from
$\fgas=0.4$ (model DCm, left panel) to $\fgas = 0.8$ (model FCm, right 
panel).   The remnants are disk-dominated and rotationally-supported, with
$\Vrot/\sigma\approx2$ at $5h^{-1}$ kpc.  The smaller system is still
massive enough to foment central star formation by driving gas inwards in
the larger galaxy, increasing the velocity dispersion by a factor of 
$\approx2$ to $\sigma \approx 80$ km s$^{-1}$ at the center each remnant.
These results are qualitatively consistent with previous work simulating
the response of galaxy disks to minor mergers at lower gas fractions 
\citep[$\fgas\approx0.1$, e.g.][]{hernquist1989a,mihos1994a,hernquist1995a}, 
but in 
our substantially
more gas-rich systems the additional supply of gas allows the larger 
galaxy
to retain comparatively more mass-weighted stellar rotation by the end of 
the simulation through continued star formation in the extended gaseous disk.
The velocity dispersion in the outer regions of our simulated remnants
increases from $\sigma \approx 30$ km s$^{-1}$ in the larger progenitor at
$5h^{-1}$ kpc before the merger to $\sigma\approx50-60$ km s$^{-1}$ 
in the remnant disk,
decreasing with progenitor gas fraction.  While the remnant properties are
not strong functions of the gas fraction for our very gas-rich mergers, with
each remnant displaying similar
rotational support and bulge-to-disk ratios ($B:D\approx1:2-1:3$, see 
Table \ref{table:remnants}), the $\Vrot/\sigma\approx2-2.5$ throughout 
most of the remnant disks substantially exceeds the 
$\Vrot/\sigma\approx1.6$ for the comparatively gas-poor mergers reported by 
\cite{bournaud2005a} for mergers of mass ratio $m_{2}:m_{1} = 8:1$ (see 
their Figure 10).  We therefore infer that increasing the gas fraction of
the disk progenitors of minor mergers beyond the maximum $\fgas = 0.2$ 
used by \cite{bournaud2005a} will improve the rotational support of 
remnant disk galaxies, but we cannot fully detail the strength of the 
effect
with the simulations presented here.
However, based upon the success of the gas-rich minor mergers in retaining
significant rotational support for the remnant disks we suggest that
gas-rich minor mergers may be an arbiter of merger-driven disk formation
and should be studied further.

\begin{figure*}
\figurenum{10}
\epsscale{1}
\plotone{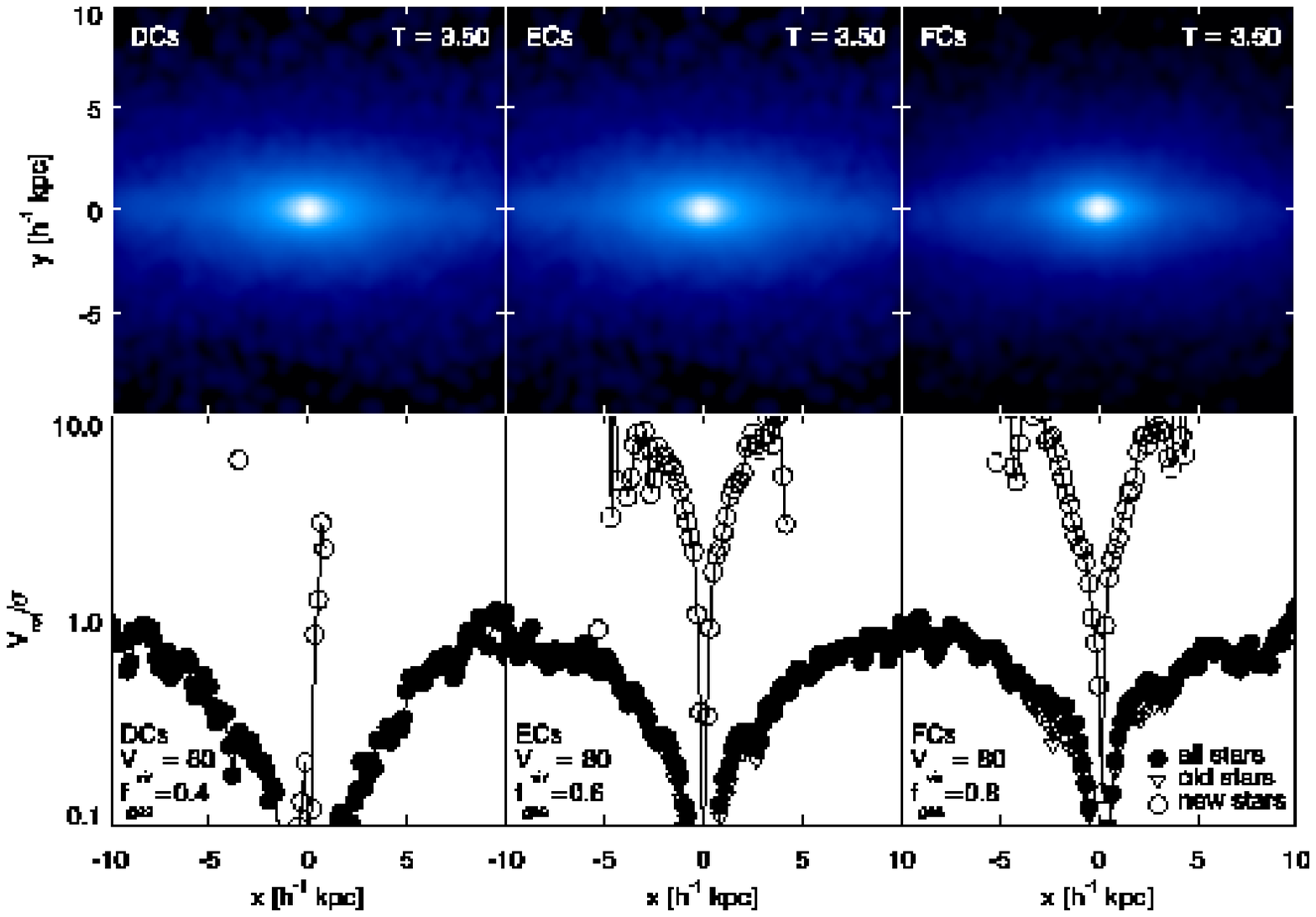}
\caption{\label{fig:vs_small}
Rotational support $\Vrot/\sigma$ and projected stellar mass
for mergers with progenitor gas fractions ranging 
from $\fgas=0.4$ (model DC, left 
panel) to $\fgas=0.8$ (model FC, right panel).
mergers for small $\Vvir=80$ km s$^{-1}$ galaxies.
Shown are the logarithmic surface density of the
stars (upper panels) and the $\Vrot/\sigma$ measured for 
all stars (solid circles), and stars formed before (open 
triangles) or after (open circles) the merger of galaxies with
moderately pressurized ISM models ($\qeos=0.5$).  The rotational
component of these dispersion dominated galaxies correlate with
the gas fraction, though the relative importance of rotation is
higher in these small mass systems than the more massive systems
in similar mergers (see Figure \ref{fig:vs_gas_fraction}).
The progenitor galaxies will likely need an initial
gas fraction of $\fgas>0.8$ and the merging system will need
$\fgas>0.5$ just prior to the final coalescence
to produce remnant disks (see Figure \ref{fig:v_feedback}). 
}
\end{figure*}

\subsection{Small Progenitor Study}
\label{subsection:results:small_mergers}

The scale-dependent physics of gas cooling and star formation may 
influence merger-driven disk formation for a mass-sequence of galaxies.  
To examine this possibility, we repeat the gas fraction study presented in
\S \ref{subsection:results:gas_fraction} substituting smaller, $\Vvir=80$
galaxies for the progenitor systems while maintaining the same orbit
(i.e. the same physical $\rperi$ and $\rinit$).
Figure \ref{fig:vs_small} shows the
$\Vrot/\sigma$ as a function of radius for mergers with moderately 
pressurized ISM models ($\qeos=0.5$) and progenitor gas fractions 
increasing from $\fgas=0.4$ (model DCs, left panel) to $\fgas=0.8$ 
(model FCs, right panel).
The remnants produced in these lower-mass
mergers are rotationally-supported only in their outer regions.
The merger with $\fgas=0.4$ progenitors (model DCs) does not contain a 
significant rotating component comprised from new stars, the only
such system in our suite of simulations.  The higher gas fraction
progenitors (models ECs and FCs) do produce rotating structures from
their young stellar components, but the average rotation of these 
systems is dominated by rotation in their large spheroid components. 

As discovered for more massive progenitors
in \S \ref{subsection:results:gas_fraction}, the gas fractions of the
merging systems both before and after the final coalescence scale 
roughly with the initial progenitor gas fraction for the same orbit
and ISM pressurization.
These systems each have gas fractions $\fgas<0.5$ just before
the final coalescence and $\fgas<0.25$ immediately afterwards, below
the gas fractions measured in mergers that produce 
remnant disks.  While the gas densities in the disks of these
systems are lower than in the more massive systems considered in
\S \ref{subsection:results:gas_fraction} and therefore produce
relatively less star formation in quiescence,  the merger timescale
for the less massive systems is longer and leads to similar 
gas fractions at the final merger.  If instead of using a physically
identical orbit for both the lower-mass and higher-mass progenitors
the orbits for the lower-mass progenitors were scaled with the disk
scale lengths, or similarly the virial radii,  the less massive
systems would have merged on a shorter timescale and may have produced
more rotationally-supported remnants.  Given the increased rotational
support in the remnants of lower-mass progenitors relative to the
higher-mass progenitors, we speculate here that lower-mass systems
may be more conducive to merger-driven disk formation than can be
inferred from the simulations we have performed.  We leave explorations
of these suspicions to future work.

\section{Discussion}
\label{section:discussion}

The suite of mergers simulated in this work indicates that
merger-driven disk formation may viable in the gas-rich mergers that
build high-redshift galaxies.  The requirements that the merging 
system is gas-dominated ($\fgas\gtrsim0.5$) during the final coalescence
to provide enough gas immediately after the merger
to sustain a remnant disk is likely a necessary but not
sufficient condition for merger-driven disk formation.  However, physical
effects that increase the gas fraction during the final merger also 
improve the structure and rotational support of remnant disks.   Increased
pressurization of the ISM,
favorable orbits with less-violent first passages or shorter merger 
timescales, small mass ratio mergers that decrease the interaction-induced
star formation in the more massive progenitors, and simply increased gas 
fractions in the progenitors all improve the rotational support of the 
remnants and in some cases lead to disk-dominated, rapidly rotating 
stellar remnants.  

The importance of gas-rich mergers for a variety of
galaxy properties has been increasingly realized
\citep[for a discussion see e.g.,][]{brook2005a}, and the 
results of our simulations of gas-rich progenitors
bear on hierarchical models for the mass 
assembly of disk galaxies.
\cite{abadi2003b} report that a substantial fraction of the
thick disk stars of a simulated disk galaxy formed in a 
cosmological setting originate as tidal debris from satellite
systems accreted after the thin disk forms.  
\cite{brook2004a,brook2005a,brook2006a}
present a different scenario where the thick disk forms
from multiple early gas-rich mergers ($\fgas\approx0.5$; C. Brook, 
private communication) and the thin disk forms later 
from the gaseous debris of these accretion events.
A recent photometric survey of the thick and thin 
components of edge-on disk galaxies by \cite{yoachim2006a}
and kinematic decompositions of the thick and thin disk 
components of two galaxies by \cite{yoachim2005a} suggest
that thick disks may be kinematically disjoint, with lower
rotational velocities than galactic thin disks.
This observational evidence supports the thick disk
accretion scenario forwarded by \cite{abadi2003b}.  However,
if substantial thick disk mass accreted after the thin disk
becomes predominately stellar then heating of the thin disk 
will occur anyway \citep[e.g.][]{quinn1993a}.
Instead, if the epoch of thick disk accretion
occurs when the large thin disk is gas-rich, then our 
simulations of minor mergers suggest that the thick disk 
may be safely accreted without unreasonably increasing the
thin disk velocity dispersion.

Our study of the formation of remnant disks in gas rich mergers 
provides a compliment to studies of elliptical galaxy formation
in gas-rich merging \citep[e.g.][]{robertson2006a}.  Major
mergers between disk galaxies with progenitor gas fractions 
$\fgas\leq0.8$ almost uniformly produce spheroid-dominated 
systems as the systems are not gas-dominated at the time of
final coalescence.  Such spheroid-dominated remnants are 
similar to our lower gas fraction models EC-FC.
As \cite{robertson2006a} demonstrate, these spheroid dominated 
systems obey elliptical galaxy scaling relations such as 
the Fundamental Plane \citep{djorgovski1987a,dressler1987a}.

Moreover, other apparently separate considerations make the
case that if ellipticals did form through mergers, the
progenitors must have been gas-rich (though not gas-dominated
as the simulations here show that these events may leave behind
systems with large disks).  For example, \cite{hernquist1993a}
showed that gas fractions larger than or of order $20-30\%$ are
required if mergers between disk galaxies are to account for the
relatively high phase space densities of local ellipticals.
Moreover, this same criterion ensures that the remnants obey
observed scaling relations \citep[e.g.][]{robertson2006a} and
match kinematic properties \citep[e.g.][]{cox2006a} of
ellipticals.

The even more gas-rich mergers with progenitor gas fractions
$\fgas > 0.8$ with disk-dominated stellar components will not
satisfy these elliptical galaxy scaling relations owing to their
rotational support.  These gas-rich merging scenarios for 
elliptical galaxy formation and remnant disk galaxy formation
are fully consistent, as the remnant stellar disks form in systems
with gas fractions large enough to sustain remnant gaseous disks
whereas elliptical galaxies are generated in highly dissipative events
where the vast majority of the gas is consumed and a large gaseous
remnant disk cannot form owing to the nearly complete depletion of gas.
In gas-rich merger events that lead to elliptical galaxy formation,
almost all of the stars form before or during the final coalescence
of the galaxies.  The production of remnant disks in mergers requires that
progenitors are comparably even more gas-rich to enable the
formation of a remnant gaseous disk
massive enough to eventually form a rapidly-rotating stellar disk
that can dominate the stellar structure of the remnant.  Naturally,
this requires that the gas fraction before the final
coalescence is $\fgas > 0.5$, in agreement with the results of 
this study.  Effects that reduce the
consumption of gas before the final coalescence assuage the 
merger-driven
production of disks,
and while our work concentrates on feedback and orbital
effects as well as the gas fraction of the progenitors 
we note that additional 
structural properties of the progenitors that reduce gas 
consumption during
the merger
(e.g. bulges) may also play a role in the formation of remnant disks.
Lastly, the range of Hubble types observed for spiral galaxies spans
a considerable range in both rotational support ($\Vrot/\sigma$) and
bulge-to-disk ratio.  The merger-driven scenario for disk galaxy
formation here may account for systems with bulge components but 
likely cannot explain very late-type systems (e.g. Sd galaxies),
which presumably must form almost entirely through dissipative
processes that do not produce hot stellar components.

We emphasize that feedback associated with black hole growth,
while essential for accounting for the observed properties
of quasars \citep{hopkins2005a,hopkins2005b,hopkins2005c} and 
the red colors
of ellipticals \citep[e.g.][]{springel2005d,hopkins2005d},
has little impact on the global structural properties of the
remnants studied here.  This is consistent with the work of 
\cite{robertson2006a} and \cite{cox2006a} who found that
black hole feedback likewise has negligible consequence for
ellipticals on scales of order the effective radius.  The
signatures of black hole growth on the stellar distribution of
merger remnants thus appears subtle and may be restricted to
modifying the characteristics of the central starburst population
formed during the merger \citep[e.g.][]{mihos1994b,mihos1994c}.

\section{Summary}
\label{section:summary}

To address the often destructive impact of the frequent merging in
the $\lcdm$ cosmology on spiral galaxies, we propose a new,
merger-driven scenario for disk galaxy formation at high redshifts
that supplements the standard picture based on dissipational
collapse \citep{white1978a,blumenthal1984a}.  In this scenario,
gas-rich mergers can form rotationally-supported gaseous structures from
residual angular momentum after the final coalescence.  This 
rapidly-rotating material will form stars, and if enough material is
available to form a substantial rotationally-supported component the
remnant can resemble a disk galaxy both structurally and kinematically.
We perform a suite of twenty-eight merger simulations to address the 
feasibility of such a scenario for disk galaxy formation and demonstrate
that remnant disks can form from sufficiently gas rich mergers with a
variety of orbits, mass ratios, and physical models for the ISM or
the inclusion of growing supermassive black holes.  We provide a
detailed summary of these results below.

\begin{itemize}
\item  We verify the results of by \cite{barnes2002a} 
and \cite{springel2005a} in demonstrating that gas-rich mergers 
can produce remnant disks.  Nearly every gas-rich merger
produces a rapidly-rotating structure from stars formed in remnant
gaseous disks after the merger.
Furthermore, we demonstrate 
that certain gas-rich mergers can produce
remnants whose entire stellar structure is rotationally-supported.
A necessary condition for the formation of these disk-dominated 
remnants is inferred to be a gas fraction of $\fgas>0.5$ just
prior to the final coalescence, which we have referred to 
as being ``gas-dominated''.
Merger-driven disk formation is shown to occur under a variety of
gas-dominated mergers and is not limited to e.g. idealized prograde-prograde 
coplanar mergers.
Favorable equal mass encounters for merger-driven disk formation 
include certain polar and inclined orbits with limited star 
formation during the first pericentric passage.  A successful
merger-driven disk formation scenario would requires extremely 
gas-rich progenitors with gas fractions $\fgas\geq0.8$, and
is therefore likely limited to high redshift galaxy assembly.

\item  The rotational support of remnants is shown to correlate
with the pressurization of the ISM.  Using the multiphase ISM
model of \cite{springel2003a} modified to allow for an
adjustable pressurization in the form of an effective equation of
state \citep{springel2005b}, the merger simulations demonstrate that
pressurized ISM models limit the amount of gas consumed before
the final coalescence of the progenitors.  The reduced star formation
increases the gaseous material available to form a rapidly-rotating
remnant disk and improves the rotational support of the remnant system.
Weakly pressurized ISM models either produce unstable progenitors owing
to the high gas fractions or have efficient star formation.  In both
cases the merging systems experience a mostly
stellar collision that simply kinematically heats the existing 
stars without producing a newly-formed disk to contribute to the 
rotation of the remnant.

\item Gas-rich minor mergers can produce disk galaxies with substantial 
rotational support ($\Vrot/\sigma>2$).  The mass ratio $m_{2}:m_{1}=8:1$
merger simulated for a range of gas fractions produces a remnant disk
galaxy rotationally-supported at all measured radii $r>1$ kpc.  The
central velocity dispersion during the merger roughly doubles and the
old stellar disk is heated.  The rotational support of the remnant
disks remaining from gas-rich minor mergers is considerably larger than that
measured for comparable gas-poor minor mergers by \cite{bournaud2005a}, who
calculate $\Vrot/\sigma\approx1.6$ for $8:1$ mass-ratio gas-poor mergers.
\end{itemize}

The simulations confirm that merger-driven disk formation from gas-rich
encounters at high redshift may be possible and highlight the role of
energetic feedback on the rotational
support of extremely gas-rich merger remnants.  
Finally, to confirm the merger-driven scenario for disk
galaxy formation as a viable supplement to the dissipational collapse
model, cosmological simulations of galaxy formation including
pressurized multiphase ISM physics
should be performed at sufficient resolution to determine
the importance of early, gas-dominated mergers to the cosmological 
frequency of disk
galaxies.  While cosmological simulations of galaxy formation with
multiphase ISM physics appear promising \citep{robertson2004a}, 
simulations with sufficient resolution to track early interactions between
gas-rich systems may provide a more comprehensive picture of
cosmological disk galaxy formation.

\acknowledgements

This work was supported in part by NSF grants ACI 96-19019, AST
00-71019, AST 02-06299, and AST 03-07690, and NASA ATP grants
NAG5-12140, NAG5-13292, and NAG5-13381.  The simulations were
performed at the Center for Parallel Astrophysical Computing at
Harvard-Smithsonian Center for Astrophysics.


\begin{thebibliography}{83}
\expandafter\ifx\csname natexlab\endcsname\relax\def\natexlab#1{#1}\fi

\bibitem[{{Abadi} {et~al.}(2003{\natexlab{a}}){Abadi}, {Navarro}, {Steinmetz},
  \& {Eke}}]{abadi2003a}
{Abadi}, M.~G., {Navarro}, J.~F., {Steinmetz}, M., \& {Eke}, V.~R.
  2003{\natexlab{a}}, \apj, 591, 499

\bibitem[{{Abadi} {et~al.}(2003{\natexlab{b}}){Abadi}, {Navarro}, {Steinmetz},
  \& {Eke}}]{abadi2003b}
---. 2003{\natexlab{b}}, \apj, 597, 21

\bibitem[{{Barnes} \& {Efstathiou}(1987)}]{barnes1987a}
{Barnes}, J., \& {Efstathiou}, G. 1987, \apj, 319, 575

\bibitem[{{Barnes}(1992)}]{barnes1992a}
{Barnes}, J.~E. 1992, \apj, 393, 484

\bibitem[{{Barnes}(2002)}]{barnes2002a}
---. 2002, \mnras, 333, 481

\bibitem[{{Barnes} \& {Hernquist}(1996)}]{barnes1996a}
{Barnes}, J.~E., \& {Hernquist}, L. 1996, \apj, 471, 115

\bibitem[{{Barnes} \& {Hernquist}(1991)}]{barnes1991a}
{Barnes}, J.~E., \& {Hernquist}, L.~E. 1991, \apj, 370, L65

\bibitem[{{Benson} {et~al.}(2004){Benson}, {Lacey}, {Frenk}, {Baugh}, \&
  {Cole}}]{benson2004a}
{Benson}, A.~J., {Lacey}, C.~G., {Frenk}, C.~S., {Baugh}, C.~M., \& {Cole}, S.
  2004, \mnras, 351, 1215

\bibitem[{{Black}(1981)}]{black1981a}
{Black}, J.~H. 1981, \mnras, 197, 553

\bibitem[{{Blumenthal} {et~al.}(1986){Blumenthal}, {Faber}, {Flores}, \&
  {Primack}}]{blumenthal1986a}
{Blumenthal}, G.~R., {Faber}, S.~M., {Flores}, R., \& {Primack}, J.~R. 1986,
  \apj, 301, 27

\bibitem[{{Blumenthal} {et~al.}(1984){Blumenthal}, {Faber}, {Primack}, \&
  {Rees}}]{blumenthal1984a}
{Blumenthal}, G.~R., {Faber}, S.~M., {Primack}, J.~R., \& {Rees}, M.~J. 1984,
  \nat, 311, 517

\bibitem[{{Bournaud} {et~al.}(2004){Bournaud}, {Combes}, \&
  {Jog}}]{bournaud2004a}
{Bournaud}, F., {Combes}, F., \& {Jog}, C.~J. 2004, \aap, 418, L27

\bibitem[{{Bournaud} {et~al.}(2005){Bournaud}, {Jog}, \&
  {Combes}}]{bournaud2005a}
{Bournaud}, F., {Jog}, C.~J., \& {Combes}, F. 2005, \aap, 437, 69

\bibitem[{{Brook} {et~al.}(2005){Brook}, {Veilleux}, {Kawata}, {Martel}, \&
  {Gibson}}]{brook2005a}
{Brook}, C., {Veilleux}, V., {Kawata}, D., {Martel}, H., \& {Gibson}, B. 2005,
  in Island Universes: Structure and Evolution of Disk Galaxies,
  astro--ph/0511002

\bibitem[{{Brook} {et~al.}(2004){Brook}, {Kawata}, {Gibson}, \&
  {Freeman}}]{brook2004a}
{Brook}, C.~B., {Kawata}, D., {Gibson}, B.~K., \& {Freeman}, K.~C. 2004, \apj,
  612, 894

\bibitem[{{Brook} {et~al.}(2006){Brook}, {Kawata}, {Gibson}, \&
  {Freeman}}]{brook2006a}
---. 2006, \apj, 639

\bibitem[{{Bullock} {et~al.}(2001){Bullock}, {Dekel}, {Kolatt}, {Kravtsov},
  {Klypin}, {Porciani}, \& {Primack}}]{bullock2001a}
{Bullock}, J.~S., {Dekel}, A., {Kolatt}, T.~S., {Kravtsov}, A.~V., {Klypin},
  A.~A., {Porciani}, C., \& {Primack}, J.~R. 2001, \apj, 555, 240

\bibitem[{{Cox} {et~al.}(2006){Cox}, , {Dutta}, {Di Matteo}, {Hernquist},
  {Hopkins}, {Robertson}, \& {Springel}}]{cox2006a}
{Cox}, T.~J., , {Dutta}, S.~N., {Di Matteo}, T., {Hernquist}, L., {Hopkins},
  P.~F., {Robertson}, B., \& {Springel}, V. 2006, \apj, submitted

\bibitem[{{de Vaucouleurs}(1948)}]{de_vaucouleurs1948a}
{de Vaucouleurs}, G. 1948, Annales d'Astrophysique, 11, 247

\bibitem[{{Di Matteo} {et~al.}(2005){Di Matteo}, {Springel}, \&
  {Hernquist}}]{di_matteo2005a}
{Di Matteo}, T., {Springel}, V., \& {Hernquist}, L. 2005, \nat, 433, 604

\bibitem[{{Djorgovski} \& {Davis}(1987)}]{djorgovski1987a}
{Djorgovski}, S., \& {Davis}, M. 1987, \apj, 313, 59

\bibitem[{{D'Onghia} \& {Burkert}(2004)}]{donghia2004a}
{D'Onghia}, E., \& {Burkert}, A. 2004, \apjl, 612, L13

\bibitem[{{Dressler} {et~al.}(1987){Dressler}, {Lynden-Bell}, {Burstein},
  {Davies}, {Faber}, {Terlevich}, \& {Wegner}}]{dressler1987a}
{Dressler}, A., {Lynden-Bell}, D., {Burstein}, D., {Davies}, R.~L., {Faber},
  S.~M., {Terlevich}, R., \& {Wegner}, G. 1987, \apj, 313, 42

\bibitem[{{Fall} \& {Efstathiou}(1980)}]{fall1980a}
{Fall}, S.~M., \& {Efstathiou}, G. 1980, \mnras, 193, 189

\bibitem[{{Font} {et~al.}(2001){Font}, {Navarro}, {Stadel}, \&
  {Quinn}}]{font2001a}
{Font}, A.~S., {Navarro}, J.~F., {Stadel}, J., \& {Quinn}, T. 2001, \apjl, 563,
  L1

\bibitem[{{Gingold} \& {Monaghan}(1977)}]{gingold1977a}
{Gingold}, R.~A., \& {Monaghan}, J.~J. 1977, \mnras, 181, 375

\bibitem[{{Governato} {et~al.}(2004){Governato}, {Mayer}, {Wadsley}, {Gardner},
  {Willman}, {Hayashi}, {Quinn}, {Stadel}, \& {Lake}}]{governato2004a}
{Governato}, F., {et~al.} 2004, \apj, 607, 688

\bibitem[{{Hernquist}(1989)}]{hernquist1989a}
{Hernquist}, L. 1989, \nat, 340, 687

\bibitem[{{Hernquist}(1990)}]{hernquist1990a}
---. 1990, \apj, 356, 359

\bibitem[{{Hernquist} \& {Barnes}(1991)}]{hernquist1991a}
{Hernquist}, L., \& {Barnes}, J.~E. 1991, \nat, 354, 210

\bibitem[{{Hernquist} \& {Mihos}(1995)}]{hernquist1995a}
{Hernquist}, L., \& {Mihos}, J.~C. 1995, \apj, 448, 41

\bibitem[{{Hernquist} {et~al.}(1993){Hernquist}, {Spergel}, \&
  {Heyl}}]{hernquist1993a}
{Hernquist}, L., {Spergel}, D.~N., \& {Heyl}, J.~S. 1993, \apj, 416, 415

\bibitem[{{Hopkins} {et~al.}(2005{\natexlab{a}}){Hopkins}, {Hernquist}, {Cox},
  {Di Matteo}, {Martini}, {Robertson}, \& {Springel}}]{hopkins2005b}
{Hopkins}, P.~F., {Hernquist}, L., {Cox}, T.~J., {Di Matteo}, T., {Martini},
  P., {Robertson}, B., \& {Springel}, V. 2005{\natexlab{a}}, \apj, 630, 705

\bibitem[{{Hopkins} {et~al.}(2005{\natexlab{b}}){Hopkins}, {Hernquist}, {Cox},
  {Di Matteo}, {Robertson}, \& {Springel}}]{hopkins2005c}
{Hopkins}, P.~F., {Hernquist}, L., {Cox}, T.~J., {Di Matteo}, T., {Robertson},
  B., \& {Springel}, V. 2005{\natexlab{b}}, ApjS, in press

\bibitem[{{Hopkins} {et~al.}(2005{\natexlab{c}}){Hopkins}, {Hernquist}, {Cox},
  {Robertson}, \& {Springel}}]{hopkins2005d}
{Hopkins}, P.~F., {Hernquist}, L., {Cox}, T.~J., {Robertson}, B., \&
  {Springel}, V. 2005{\natexlab{c}}, \apj, in press

\bibitem[{{Hopkins} {et~al.}(2005{\natexlab{d}}){Hopkins}, {Hernquist},
  {Martini}, {Cox}, {Robertson}, {Di Matteo}, \& {Springel}}]{hopkins2005a}
{Hopkins}, P.~F., {Hernquist}, L., {Martini}, P., {Cox}, T.~J., {Robertson},
  B., {Di Matteo}, T., \& {Springel}, V. 2005{\natexlab{d}}, \apjl, 625, L71

\bibitem[{{Hultman} \& {Pharasyn}(1999)}]{hultman1999a}
{Hultman}, J., \& {Pharasyn}, A. 1999, \aap, 347, 769

\bibitem[{{Katz} \& {Gunn}(1991)}]{katz1991a}
{Katz}, N., \& {Gunn}, J.~E. 1991, \apj, 377, 365

\bibitem[{{Katz} {et~al.}(1992){Katz}, {Hernquist}, \& {Weinberg}}]{katz1992a}
{Katz}, N., {Hernquist}, L., \& {Weinberg}, D.~H. 1992, \apjl, 399, L109

\bibitem[{{Kennicutt}(1998)}]{kennicutt1998a}
{Kennicutt}, R.~C. 1998, \apj, 498, 541

\bibitem[{{Lacey} \& {Cole}(1993)}]{lacey1993a}
{Lacey}, C., \& {Cole}, S. 1993, \mnras, 262, 627

\bibitem[{{Lucy}(1977)}]{lucy1977a}
{Lucy}, L.~B. 1977, \aj, 82, 1013

\bibitem[{{Maller} \& {Dekel}(2002)}]{maller2002a}
{Maller}, A.~H., \& {Dekel}, A. 2002, \mnras, 335, 487

\bibitem[{{McKee} \& {Ostriker}(1977)}]{mckee1977a}
{McKee}, C.~F., \& {Ostriker}, J.~P. 1977, \apj, 218, 148

\bibitem[{{Mihos} \& {Hernquist}(1994{\natexlab{a}})}]{mihos1994b}
{Mihos}, J.~C., \& {Hernquist}, L. 1994{\natexlab{a}}, \apj, 427, 112

\bibitem[{{Mihos} \& {Hernquist}(1994{\natexlab{b}})}]{mihos1994a}
---. 1994{\natexlab{b}}, \apjl, 425, L13

\bibitem[{{Mihos} \& {Hernquist}(1994{\natexlab{c}})}]{mihos1994c}
---. 1994{\natexlab{c}}, \apjl, 431, L9

\bibitem[{{Mihos} \& {Hernquist}(1996)}]{mihos1996a}
---. 1996, \apj, 464, 641

\bibitem[{{Mo} {et~al.}(1998){Mo}, {Mao}, \& {White}}]{mo1998a}
{Mo}, H.~J., {Mao}, S., \& {White}, S.~D.~M. 1998, \mnras, 295, 319

\bibitem[{{Navarro} {et~al.}(1997){Navarro}, {Frenk}, \&
  {White}}]{navarro1997a}
{Navarro}, J.~F., {Frenk}, C.~S., \& {White}, S.~D.~M. 1997, \apj, 490, 493

\bibitem[{{Navarro} \& {Steinmetz}(2000)}]{navarro2000a}
{Navarro}, J.~F., \& {Steinmetz}, M. 2000, \apj, 538

\bibitem[{{Navarro} \& {White}(1994)}]{navarro1994a}
{Navarro}, J.~F., \& {White}, S.~D.~M. 1994, \mnras, 267, 401

\bibitem[{{Okamoto} {et~al.}(2005){Okamoto}, {Eke}, {Frenk}, \&
  {Jenkins}}]{okamoto2005a}
{Okamoto}, T., {Eke}, V.~R., {Frenk}, C.~S., \& {Jenkins}, A. 2005, \mnras,
  363, 1299

\bibitem[{{Peebles}(1969)}]{peebles1969a}
{Peebles}, P.~J.~E. 1969, \apj, 155, 393

\bibitem[{{Quinn} {et~al.}(1993){Quinn}, {Hernquist}, \&
  {Fullagar}}]{quinn1993a}
{Quinn}, P.~J., {Hernquist}, L., \& {Fullagar}, D.~P. 1993, \apj, 403, 74

\bibitem[{{Robertson} {et~al.}(2006){Robertson}, {Cox}, {Hernquist}, {Franx},
  {Hopkins}, {Martini}, \& {Springel}}]{robertson2006a}
{Robertson}, B., {Cox}, T.~J., {Hernquist}, L., {Franx}, M., {Hopkins}, P.~F.,
  {Martini}, P., \& {Springel}, V. 2006, \apj, 641

\bibitem[{{Robertson} {et~al.}(2004){Robertson}, {Yoshida}, {Springel}, \&
  {Hernquist}}]{robertson2004a}
{Robertson}, B., {Yoshida}, N., {Springel}, V., \& {Hernquist}, L. 2004, \apj,
  606, 32

\bibitem[{{Sommer-Larsen} {et~al.}(2003){Sommer-Larsen}, {G{\" o}tz}, \&
  {Portinari}}]{sommer-larsen2003a}
{Sommer-Larsen}, J., {G{\" o}tz}, M., \& {Portinari}, L. 2003, \apj, 596, 47

\bibitem[{{Sommer-Larsen} {et~al.}(1999){Sommer-Larsen}, {Gelato}, \&
  {Vedel}}]{sommer-larsen1999a}
{Sommer-Larsen}, J., {Gelato}, S., \& {Vedel}, H. 1999, \apj, 519, 501

\bibitem[{{Springel} {et~al.}(2005{\natexlab{a}}){Springel}, {Di Matteo}, \&
  {Hernquist}}]{springel2005d}
{Springel}, V., {Di Matteo}, T., \& {Hernquist}, L. 2005{\natexlab{a}}, \apjl,
  620, L79

\bibitem[{{Springel} {et~al.}(2005{\natexlab{b}}){Springel}, {Di Matteo}, \&
  {Hernquist}}]{springel2005b}
---. 2005{\natexlab{b}}, \mnras, 361, 776

\bibitem[{{Springel} \& {Hernquist}(2003)}]{springel2003a}
{Springel}, V., \& {Hernquist}, L. 2003, \mnras, 339, 289

\bibitem[{{Springel} \& {Hernquist}(2005)}]{springel2005a}
---. 2005, \apjl, 622, 9

\bibitem[{{Steinmetz} \& {Muller}(1995)}]{steinmetz1995a}
{Steinmetz}, M., \& {Muller}, E. 1995, \mnras, 276, 549

\bibitem[{{Thacker} \& {Couchman}(2000)}]{thacker2000a}
{Thacker}, R.~J., \& {Couchman}, H.~M.~P. 2000, \apj, 545, 728

\bibitem[{{Thacker} \& {Couchman}(2001)}]{thacker2001a}
---. 2001, \apjl, 555, L17

\bibitem[{{Thakar} \& {Ryden}(1996)}]{thakar1996a}
{Thakar}, A.~R., \& {Ryden}, B.~S. 1996, \apj, 461, 55

\bibitem[{{Thakar} \& {Ryden}(1998)}]{thakar1998a}
---. 1998, \apj, 506, 93

\bibitem[{{Toomre}(1964)}]{toomre1964a}
{Toomre}, A. 1964, \apj, 139, 1217

\bibitem[{{Toomre}(1977)}]{toomre1977a}
{Toomre}, A. 1977, in Evolution of Galaxies and Stellar Populations, 401--+

\bibitem[{{Toth} \& {Ostriker}(1992)}]{toth1992a}
{Toth}, G., \& {Ostriker}, J.~P. 1992, \apj, 389, 5

\bibitem[{{van den Bosch}(1998)}]{van_den_bosch1998a}
{van den Bosch}, F.~C. 1998, \apj, 507, 601

\bibitem[{{van den Bosch}(2001)}]{van_den_bosch2001b}
---. 2001, \mnras, 327, 1334

\bibitem[{{van den Bosch} {et~al.}(2002){van den Bosch}, {Abel}, {Croft},
  {Hernquist}, \& {White}}]{van_den_bosch2002a}
{van den Bosch}, F.~C., {Abel}, T., {Croft}, R.~A.~C., {Hernquist}, L., \&
  {White}, S.~D.~M. 2002, \apj, 576, 21

\bibitem[{{van den Bosch} {et~al.}(2001){van den Bosch}, {Burkert}, \&
  {Swaters}}]{van_den_bosch2001a}
{van den Bosch}, F.~C., {Burkert}, A., \& {Swaters}, R.~A. 2001, \mnras, 326,
  1205

\bibitem[{{Velazquez} \& {White}(1999)}]{velazquez1999a}
{Velazquez}, H., \& {White}, S.~D.~M. 1999, \mnras, 304, 254

\bibitem[{{Vitvitska} {et~al.}(2002){Vitvitska}, {Klypin}, {Kravtsov},
  {Wechsler}, {Primack}, \& {Bullock}}]{vitvitska2002a}
{Vitvitska}, M., {Klypin}, A.~A., {Kravtsov}, A.~V., {Wechsler}, R.~H.,
  {Primack}, J.~R., \& {Bullock}, J.~S. 2002, \apj, 581, 799

\bibitem[{{Walker} {et~al.}(1996){Walker}, {Mihos}, \&
  {Hernquist}}]{walker1996a}
{Walker}, I.~R., {Mihos}, J.~C., \& {Hernquist}, L. 1996, \apj, 460, 121

\bibitem[{{Weil} {et~al.}(1998){Weil}, {Eke}, \& {Efstathiou}}]{weil1998a}
{Weil}, M.~L., {Eke}, V.~R., \& {Efstathiou}, G. 1998, \mnras, 300, 773

\bibitem[{{White} \& {Rees}(1978)}]{white1978a}
{White}, S.~D.~M., \& {Rees}, M.~J. 1978, \mnras, 183, 341

\bibitem[{{Yepes} {et~al.}(1997){Yepes}, {Kates}, {Khokhlov}, \&
  {Klypin}}]{yepes1997a}
{Yepes}, G., {Kates}, R., {Khokhlov}, A., \& {Klypin}, A. 1997, \mnras, 284,
  235

\bibitem[{{Yoachim} \& {Dalcanton}(2005)}]{yoachim2005a}
{Yoachim}, P., \& {Dalcanton}, J.~J. 2005, \apj, 624, 701

\bibitem[{{Yoachim} \& {Dalcanton}(2006)}]{yoachim2006a}
---. 2006, \aj, 131, 226

\end{thebibliography}
\end{document}